\documentclass{emulateapj}

\usepackage{verbatim}
\usepackage{natbib}
\usepackage{amsmath}
\usepackage{amsbsy}
\slugcomment{Draft}
\shorttitle{Proximity Effect and Mass of Quasar Host Halos}
\shortauthors{Faucher-Gigu\`ere et al.}

\newcommand{\Lya}{\mbox{Ly$\alpha$}}
\newcommand{\Gbkg}{\mbox{$\Gamma^{bkg}$}}
\newcommand{\Msun}{\mbox{M$_{\odot}$}}
\newcommand{\bs}[1]{{\bf{#1}}}
\newcommand{\transpose}{\mbox{${}^{\textrm{T}}$}}

\begin{document}

\title{The Line-of-Sight Proximity Effect and the Mass of Quasar Host Halos}

\author{Claude-Andr\'e Faucher-Gigu\`ere\altaffilmark{1}, Adam Lidz\altaffilmark{1}, Matias Zaldarriaga\altaffilmark{1,2}, Lars Hernquist\altaffilmark{1}}
\altaffiltext{1}{Department of Astronomy, Harvard University, Cambridge, MA, 02138, USA; cgiguere@cfa.harvard.edu.}
\altaffiltext{2}{Jefferson Physical Laboratory, Harvard University, Cambridge, MA, 02138, USA.}

\begin{abstract}
We show that the Lyman-$\alpha$ (\Lya) optical depth statistics in the
proximity regions of quasar spectra depend on the mass of
the dark matter halos hosting the quasars.
This is owing to both the overdensity around the quasars and the associated infall of
gas toward them. 
For a fiducial quasar host halo mass of \mbox{$3.0\pm1.6~h^{-1}\times10^{12}$~\Msun}, as inferred by Croom et al. from clustering in the 2dF QSO Redshift Survey, we show that estimates of the ionizing background (\Gbkg) from proximity effect measurements could be biased high by a factor of $\approx2.5$ at $z=3$ owing to neglecting these effects alone. 
The clustering of galaxies and other active galactic nuclei around the proximity effect quasars enhances the local background, but is not expected to skew measurements by more than a few percent.
Assuming the measurements of \Gbkg~based on the mean flux decrement in the \Lya~forest to be free of bias, we demonstrate how the proximity effect analysis can be inverted to measure the mass of the dark matter halos hosting quasars.
In ideal conditions, such a measurement could be made with a precision comparable to the best clustering constraints to date from a modest sample of only about 100 spectra.
We discuss observational difficulties, including continuum flux estimation, quasar systematic redshift determination, and quasar variability, which make accurate proximity effect measurements challenging in practice.
These are also likely to contribute to the discrepancies between existing proximity effect and flux decrement measurements of \Gbkg.
\end{abstract}


\section{INTRODUCTION}
\label{introduction}
\begin{figure*}[ht]
\begin{center}
\includegraphics[width=0.55\textwidth,angle=-90]{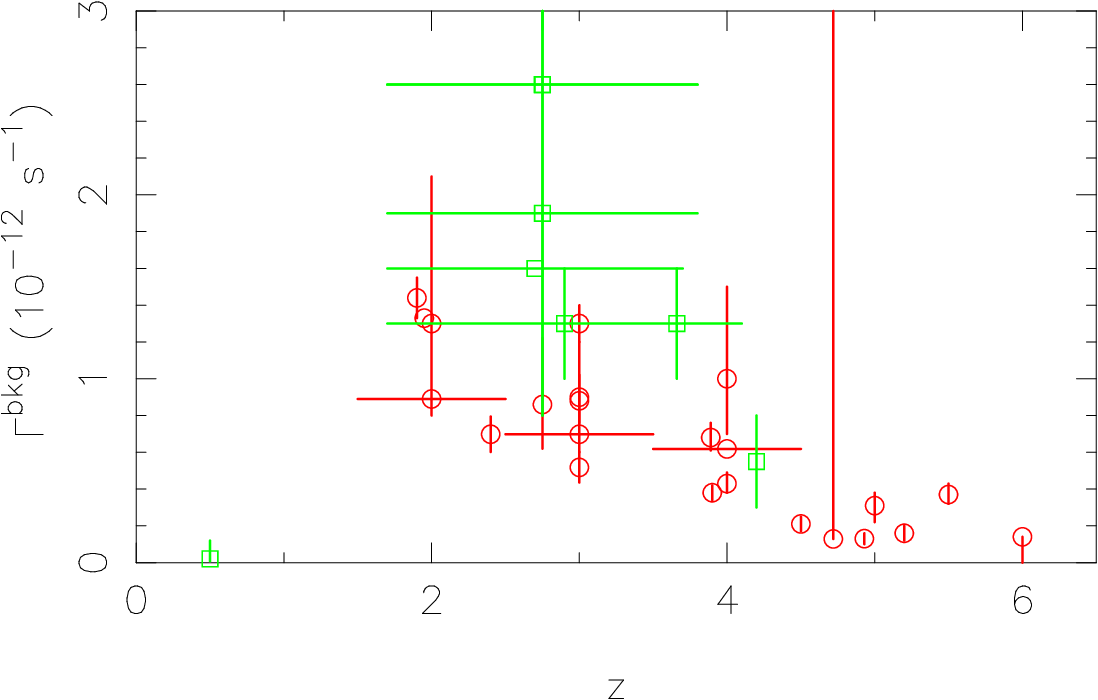}
\end{center}
\caption{Existing measurements of \Gbkg~using the flux decrement (red, circles) and the proximity effect methods (green, squares).
The horizontal error bars, where present, indicate the redshift range over which the measurement applies.
The vertical bars show the reported uncertainties.
The early proximity effect estimate \mbox{$\Gamma^{bkg}\gtrsim7.8\times10^{12}$ s$^{-1}$} at $z=3.75$ of \cite{1987ApJ...319..709C} is not shown.
In the redshift range where measurement have been made using both methods, the proximity effect estimates are generally higher by a factor of $2-3$.
The scatter between the different proximity effect measurements is also larger than in the flux decrement case.
Numerical values, references, and additional comments are provided in Table \ref{existing measurements table}.
}
\label{existing measurements figure}
\end{figure*}

As the geometrical properties and initial conditions of the Universe
are becoming fairly well known \citep[e.g.,][]{2006astro.ph..3449S}, an
increasingly important aspect of cosmology concerns the emergence and
evolution of complex, non-linear structures. 
Light sources --- stellar and quasistellar -- are tracers of these structures and it is thus of prime interest to study their nature and evolution.
A measure of the stellar and quasistellar activity is given by the
amount and distribution of photons with energy above the hydrogen
ionization potential in the intergalactic medium (IGM). 
The relevant integral quantity is the redshift ($z$)-dependent background hydrogen photoionization rate, $\Gamma^{bkg}(z)$, which measures the integrated contributions from all sources of ionizing photons.
Several measurements of $\Gamma^{bkg}$ exist in the literature (see Figure \ref{existing measurements figure}, complemented by the numerical values tabulated in Table \ref{existing measurements table} of Appendix \ref{measurements appendix}). 
Unfortunately, the agreement between various studies using different methods is poor.

Two main methods have been used to estimate $\Gamma^{bkg}$: the
flux decrement method and the line-of-sight proximity effect method.
In the flux decrement method \citep{1997ApJ...489....7R, 1999ApJ...525L...5S, 2001ApJ...549L..11M, 2003MNRAS.342.1205M, 2004ApJ...617....1T, 2005MNRAS.357.1178B, 2005MNRAS.360.1373K, 2005MNRAS.361...70J}, one considers the portion of the \mbox{Lyman$-\alpha$} (\Lya) forest along lines of sight to quasars (QSOs) away from the latter.
The parameter $\mu \propto \Omega_{b}^{2}h^{3}/\Gamma^{bkg}$, where $\Omega_{b}$ is the cosmic baryon density in units of the critical density and \mbox{$h=H_{0} / 100$ km s$^{-1}$ Mpc$^{-1}$}, is obtained by requiring the mean flux decrement (\mbox{$D\equiv \langle 1-e^{-\tau} \rangle$}) in a cosmological simulation to agree with the mean decrement observed in quasar spectra.
In the proximity effect method (\citealt{1987ApJ...319..709C, 1988ApJ...327..570B, 1991ApJ...367...19L, 1993ApJ...413L..63K, 1994ApJ...428..574W, 1995MNRAS.273.1016C, 1996ApJ...466...46G, 1996MNRAS.280..767S, 2000ApJS..130...67S}), in which we will be most interested here, one focuses on the ``proximity region'' near the quasar, where the ionizing flux emitted by the quasar itself is comparable to the background. (For related work on the {\em transverse} proximity effect
see, e.g., \citealt{2004ApJ...610..105S, 2004ApJ...610..642C, 2004ApJ...612..706A, 2006astro.ph..6084H}.)

In the simplest proximity effect model, the quasar lies in a random region of the Universe
and the hydrogen gas is in photoionization equilibrium. Neglecting all motion of the gas,
the \Lya~optical depth at any point is then given by
\begin{equation}
\label{simple tau scaling}
\tau^{prox}
=
\frac{
\tau^{off}
}
{
1 + \omega(r)
}
,
\end{equation}
where $\omega(r) \equiv \Gamma^{QSO}(r)/\Gamma^{bkg}$, $\Gamma^{QSO}(r)$ is the contribution to the photoionization rate owing to the quasar itself at proper distance $r$ from it, and $\tau^{off}$
is the optical depth that would be obtained if the quasar were turned off.
Near the quasar, \mbox{$\Gamma^{QSO} \propto r^{-2}$} (and hence $\omega$) is large, causing a statistical decrease in observed \Lya~absorption. 
Authors have generally fitted a model, introduced by \cite{1988ApJ...327..570B} (hereafter BDO), for the variation of the density of absorption lines near the quasars from spectra to estimate $\Gamma^{bkg}$.
Integration over known sources of radiation \citep[e.g.,][]{2001ApJ...546..665S, 2006astro.ph..5678H} can also be used to estimate \Gbkg.
However, this method provides only lower bounds and much of the interest in radiation backgrounds lies in determining whether the measured values agree with the summed contribution of resolved sources.
Thus, \Gbkg~should be measured independently of such estimates and we will not consider them further in this paper.

The proximity effect method has generally yielded measurements of \Gbkg~higher than flux decrement estimates (Figure \ref{existing measurements figure}), suggesting unmodeled biases in one or both methods.
There are indeed potential sources of bias in the proximity effect method, mostly owing to the peculiar
environments in which quasars are thought to reside, that have not generally been
modeled in proximity effect analyses (although see \citealt{1995ApJ...448...17L}, \citealt{2005MNRAS.361.1015R}, \citealt{2007MNRAS.tmp..238G}, and \S \ref{previous work} for a comparison with previous work). These potential biases are a focus of this paper. 

In cold dark matter (CDM) models of structure formation, quasars are expected to preferentially form in overdense regions of the universe \citep[e.g.,][]{2005Natur.435..629S}, violating the assumption that they lie in random
locations.
The overdense environments of quasars not only affect the local density of the absorbing gas, but also its peculiar velocity field.
General infall of the gas toward the density peaks where quasars reside is expected \citep{1995ApJ...448...17L, 2004MNRAS.347...59B} and in fact may have been directly detected \citep{2003Natur.421..341B}.
Moreover, galaxies and other active galactic nuclei (AGN) should preferentially form in overdense regions \citep{1991ApJ...379..440B} and therefore cluster around quasars.
The local ``background'' flux near the quasar may be more intense than average.
Thermal broadening owing to the finite temperature of the gas also causes redshift-space distortions, increasing the equivalent width of absorption lines.

The flux decrement method is also not guaranteed to be unbiased, most
importantly as it requires estimation of the unabsorbed continuum
level.
This is difficult to do reliably, especially at high redshifts
where unabsorbed portions become increasingly rare, if existent at all, in
quasar spectra. Alternatively, one can estimate the unabsorbed continuum
level by extrapolating from redward of Ly-$\alpha$ (e.g., \citealt{1993ApJ...418..585P}).
However, there may be a break in the quasar continuum close to Ly-$\alpha$ 
\citep{1997ApJ...475..469Z, 2002ApJ...565..773T}, which could bias resulting
estimates of the flux decrement \citep{2003MNRAS.342L..79S}. For further discussion
of flux decrement estimation and related issues see, e.g., \citealt{2003MNRAS.342L..79S}, \citealt{2004AJ....128.1058T}, and \citealt{2006ApJ...638...27L}.

The flux decrement method is also sensitive to assumed cosmological
parameters, most notably $\Omega_{b}$,
through the factor $\Omega_{b} h^{2}$ in $\mu$.
If an independent measurement of \Gbkg~is available, then the flux decrement may be used to infer $\Omega_{b}$.
The proximity effect provides such a measurement and,
as integration over known sources can only provide lower bounds on
\Gbkg, is the only known way to use the
flux decrement to obtain a full measurement of $\Omega_{b}$.

Integration over sources from a luminosity at rest-frame frequencies below the Lyman limit requires knowledge of the fraction of ionizing photons which escape the immediate environment of the emitters and actually have an effect on the surrounding IGM.
As recently shown by \cite{2006astro.ph..6635S}, the escape fraction
of Lyman break galaxies (LBG) at $z\sim3$, whose contribution may well
dominate \Gbkg~at these redshifts \citep{2001ApJ...546..665S}, is
highly uncertain.
Agreement between measurements of $\Omega_{b}$ from light element abundances interpreted in the context of Big Bang nucleosynthesis \citep[e.g.,][]{2001ApJ...552L...1B}, cosmic microwave background anisotropies \citep[e.g.,][]{2006astro.ph..3449S}, and the flux decrement would provide a powerful consistency check of the standard cosmological model as well as of the gravitational paradigm for the \Lya~forest \citep{2002ApJ...564..525H}.

A detailed understanding of the proximity effect would also provide insight into the astrophysics of quasar environments and in turn could be used to study these.
For example, if the dark matter (DM) halos in which quasars lie can be
shown to induce significant biases in measurements of \Gbkg, then one
could parameterize these biases with respect to the quasar host
halo mass, $M_{DM}$.
Knowledge of the quasar host halo mass and constraints on its dependence on quasar luminosity would be tremendously useful in determining how activity
is triggered in AGN.
This is especially true in a picture in which activity is related to mergers \citep[e.g.,][]{2006ApJS..163....1H,
2006ApJ...641...41L}, as such events inevitably increase the host halo mass.

Recent clustering measurements suggest a universal quasar host halo mass. 
For example, \cite{2005MNRAS.356..415C} find that quasars lie in dark
matter halos of mass \mbox{$3.0\pm1.6~h^{-1}\times10^{12}$~\Msun} based on an
analysis of the quasar two-point correlation function as measured from
2dF QSO Redshift Survey (2QZ) quasars, only weakly dependent on redshift and
with no evidence for a luminosity dependence.
If confirmed, this result would provide constraints on quasar
evolution and call for an explanation of the physical origin of this
critical mass for nuclear activity.
Given that halos continuously grow through mergers, a universal mass
for AGN activity also constrains quasar lifetimes.
Combined with a relation between the mass of the central black hole ($M_{BH}$) and the mass of its host halo \citep[e.g.,][]{2002ApJ...578...90F}, a 
universal halo mass would also suggest a universal black hole mass, which may be at odds with
the variety of nuclear black hole masses observed in the local
universe \citep[e.g.,][]{1995ARA&A..33..581K} if most galactic nuclei
have gone through an active phase, although it is possible that the
$M_{BH}-M_{DM}$ relation evolves with redshift (Hopkins et al. 2006c\nocite{2006astro.ph.11792H}, Hopkins et al., in preparation).
If the universal quasar host halo mass is correct, it would provide
evidence for a redshift-dependent $M_{BH}-M_{DM}$ relation.
As practically all the evidence for the quasar host halo mass universality is inferred from clustering measurements, independent quasar host halo mass estimates
are clearly desirable. 

In this work, we quantify the effects of quasar host halos on the \Lya~statistics in the proximity regions of quasars and the biases they induce in proximity effect measurements of \Gbkg~(\S \ref{overdensities infall and clustering} and \S \ref{proximity effect biases}), develop a Monte Carlo-based method for making unbiased \Gbkg~measurements (\S \ref{unbiased llhd}), and demonstrate how the analysis can be inverted to use the proximity effect to measure the masses of dark matter halos hosting quasars (\S \ref{halo mass}).
Throughout most of the paper, we assume that the analyses are performed in ideal quasar spectra.
In \S \ref{practical considerations}, we consider observational difficulties which make accurate proximity effect measurements challenging in practice and which may also contribute to the discrepancies between proximity effect and flux decrement measurements of \Gbkg.
We compare with previous work in \S \ref{previous work} and
conclude in \S \ref{conclusion}.
We begin by summarizing conventions used throughout the paper in
\S \ref{conventions} and describing our numerical framework in
\S \ref{numerical simulations}.

Most previous studies of the proximity effect 
were based on a picture of the \Lya~forest as arising from absorption by discrete gas clouds and on counting absorption lines.
Here, we adopt the modern picture of the forest as arising from absorption by smooth density fluctuations imposed on the warm photoionized IGM as a natural consequence of hierarchical structure formation within CDM models.
This picture, based on a combination of detailed hydrodynamical
simulations \citep{1994ApJ...437L...9C, 1995ApJ...453L..57Z, 1996ApJ...457L..51H, 1996ApJ...457L..57K, 1996ApJ...471..582M, 1998MNRAS.301..478T, 1999ApJ...511..521D} and analyses of high-resolution, high signal-to-noise quasar spectra \citep{1996ApJ...472..509L, 1997ApJ...484..672K, 2002MNRAS.335..555K} with matching properties, is strongly supported by all available evidence.

As we were finalizing this work we became aware of a related study by \cite{2007astro.ph..1012K}. These authors also consider constraining the mass of quasar host halos from 
the \Lya~forest. They, however, focus on using quasar pairs and work under the hypothesis (supported by observations) that the \emph{transverse} proximity effect is negligible.
We, on the other hand, focus on the {\em line-of-sight} proximity effect and model it in detail. 
The two works, while broadly consistent where there is overlap, are hence complementary.

\section{CONVENTIONS}
\label{conventions}
In this paper, $\widehat{X}$ denotes a random variable, $X$ a
realized value, and $f_{\widehat{X}}(X)$ the probability density
function (PDF) of $\widehat{X}$ evaluated at $X$.
The linear overdensity $\delta$ is defined such that $\rho / \bar{\rho} = 1 + \delta$, where $\rho$ is the local mass density and $\bar{\rho}$ is the mean density of the Universe.
We assume a flat $\Lambda CDM$ cosmology similar to the one inferred
from \emph{Wilkinson Microwave Anisotropy Probe} (WMAP) measurements
of the cosmic microwave background \citep{2006astro.ph..3449S}; the exact parameters assumed in
our cosmological simulations are given in Table \ref{sim params}.
We select ``typical" quasars to have spectral indexes ($L_{\nu} \propto
\nu^{-\alpha}$) $\alpha=1.57$ blueward of \Lya.
For the typical specific luminosity at the Lyman limit, we take the median $\log_{10}(L_{912} / \textrm{erg s$^{-1}$ cm$^{-2}$ Hz$^{-1}$})=31.1$ for the quasars in the proximity effect analysis of \cite{2000ApJS..130...67S}.
When we refer to quasars with luminosity one standard deviation from typical, we approximate (generously) the standard deviation in \cite{2000ApJS..130...67S} to $\sigma_{\log_{10}{L_{912}}}=1$.

\section{NUMERICAL SIMULATIONS AND MOCK SPECTRA}
\label{numerical simulations}
To study the effects of various potential sources of bias on
the measurement of \Gbkg~from the proximity effect, we will make use
of mock quasar spectra generated from cosmological simulations.
In generating the mock spectra, we will turn on and off different
effects, examining their impact on the derived \Gbkg.
In this section, we describe the simulations that we use and how basic mock
spectra are constructed from them.
The reader who is not concerned with such numerical details is
encouraged to skip to \S \ref{overdensities infall and clustering}, where we begin our discussion of the effects of overdensities and redshift-space distortions. 

\subsection{Cosmological Dark Matter Simulations}
The first step in our modeling of the \Lya~forest is to compute the
dark matter density and velocity fields in a simulation box at
different redshifts.
We do so using the GADGET$-$2 \citep{2005MNRAS.364.1105S}
cosmological code.
The simulation parameters are given in Table \ref{sim params}.
When generating density and velocity profiles and constructing optical
depth PDFs in the following sections, significant numbers of halos in
different mass ranges are required.
This motivates our choice of a large simulation box, with side length
$100~h^{-1}$ comoving Mpc.
In order to compare our results with existing proximity effect
analyses, it is also necessary to run the simulation to redshifts at least as low as $z\approx2$.
We use $N$-body outputs with $256^{3}$ particles at $z=2, 3,~\textrm{and}~4$; the 
least massive resolved halos have mass \mbox{$4.96~h^{-1}\times10^{9}$ \Msun}.
Ideally, for a detailed comparison with Ly$\alpha$ forest data, we would use a higher
resolution, fully hydrodynamic simulation (e.g., \citealt{2006MNRAS.367.1655V}). Our
present $N$-body approach should, however, be adequate for investigating the {\em relative} impact of quasar environments on Ly$\alpha$ absorption statistics.
\begin{center}
\begin{deluxetable}{cc}
\tablecaption{Simulation Parameters\label{sim params}}
\tablewidth{0pt}

\tablehead{
\multicolumn{2}{c}{$N-$body Dark Matter Simulation} \\
\hline
\colhead{Parameter} & \colhead{Value}
}

\startdata
Code                & GADGET-2\tablenotemark{a}         \\
Box Side Length     & 100 $h^{-1}$ comoving Mpc         \\
Boundary Conditions & Periodic                          \\
Number of Particles & $256^{3}$                         \\
Particle Mass       & $4.96 \times 10^{9}~h^{-1}$ \Msun \\
$\Omega_{m}$        & 0.3                               \\
$\Omega_{\Lambda}$  & 0.7                               \\
$\Omega_{b}$        & 0.04                              \\
$h$                 & 0.7                               \\
$\sigma_{8}$        & 0.9                               \\
$n_{s}$             & 1                                 \\

\hline
\multicolumn{2}{c}{Friends-of-Friends Halo Finder} \\
\hline

Linking Length              & 0.2 $\langle \Delta r_{p} \rangle$\tablenotemark{b} \\
Minimum Number of Particles & 31             \\

\hline
\multicolumn{2}{c}{Grid} \\
\hline

Grid Interpolation    & CIC on $256^{3}$ grid points \\
Line of Sight Interpolation & TSC

\enddata

\tablenotetext{a}{\cite{2005MNRAS.364.1105S}.}
\tablenotetext{b}{Mean interparticle spacing.}

\end{deluxetable}
\end{center}

\subsection{Mock Quasar Spectra from Dark Matter Simulations}
\label{mock spectra}
In the simplest model, the neutral hydrogen (HI) density at any given
point in the Universe is given by photoionization equilibrium.
Let $\Gamma^{tot}$ (\mbox{s$^{-1}$}) be the total HI photoionization
rate per atom, $R(T)$ (\mbox{cm$^{3}$ s$^{-1}$}) the recombination rate, $n_{\textrm{HI}}$ the HI number density, $n_{\textrm{HII}}$ the ionized hydrogen (HII) number density, and $n_{e}$ be the free electron number density in proper units.
Then equilibrium requires
\begin{equation}
\Gamma^{tot} n_{\textrm{HI}}
=
R(T)
n_{e}
n_{\textrm{HII}}.
\end{equation}
For a small ionized fraction $x_{i} \equiv
n_{\textrm{HII}}/n_{\textrm{HI}} \ll 1$ (which is certainly true in
the IGM at the redshifts of interest here, $z \lesssim 6$, as
indicated by the absence of a Gunn-Peterson\nocite{1965ApJ...142.1633G} in
the known quasars at these redshifts),
\begin{equation}
\label{hi number density}
n_{\textrm{HI}} =
\frac{R(T) n_{tot}^{2}}
{\Gamma^{tot}},
\end{equation}
where $n_{tot}=n_{\textrm{HI}} + n_{\textrm{HII}} \approx \Omega_{b} X_{H} \rho_{c}/m_{p}$ is the total hydrogen number density.
We take the cosmic hydrogen mass fraction $X_{H}=0.75$ \citep[e.g.,][]{1999PhRvL..82.4176B}.
The above expression for $n_{\textrm{HI}}$ neglects the helium
contribution to the free electron density.
For fully singly ionized helium, we make an error of 8\% on $n_{e}$
and hence on $n_{\textrm{HI}}$; for fully doubly ionized helium, the error is doubled.
Simplifying approximations such as this one have no impact on our discussion,
which should be relatively independent of the details of the
cosmology.
For the recombination rate, we use the approximate expression
\begin{equation}
\label{H recombination coefficient}
R(T) =
4.2 \times 10^{-13}
\left(
\frac{T}
{10^{4}\textrm{ K}}
\right)^{-0.7}
\textrm{ cm}^{3}\textrm{ s}^{-1},
\end{equation}
\citep{1997MNRAS.292...27H} where $T$ is the local gas temperature.

For $\delta \lesssim 5$, \cite{1997MNRAS.292...27H} derived the
equation of state
\begin{equation}
\label{igm eos}
T = T_{0}(1 + \delta)^{\beta},
\end{equation}
for the IGM, where $T_{0}(z)$ is the temperature of a fluid element which remains at the cosmic mean density and $\beta(z)$ parameterizes the density dependence.
Detailed expressions for $T_{0}$ and $\beta$, which we use in our numerical
calculations, are given in Appendix \ref{accurate expressions}.
To order of magnitude, \mbox{$T_{0} \sim 10^{4}$ K} and in the limit of early reionization ($z_{reion} \gtrsim 10$, consistent with recent measurements of the Thomson optical depth to reionization by Page et al. 2006\nocite{2006astro.ph..3450P}), $\beta \to 0.62$. Temperature measurements from the $z \sim 3$ Ly$\alpha$ forest favor slightly
larger temperatures and a slightly less steep density dependence (e.g., \citealt{2001ApJ...562...52M}), possibly
owing to recent HeII reionization. Our basic conclusions should, however, not depend on the precise 
temperature-density relation assumed. 

For a quasar with isotropic specific luminosity $L_{\nu}=L_{Q}(\nu / \nu_{Ly})^{-\alpha_{Q}}$, where $\nu_{Ly}$ is the hydrogen ionization frequency in its ground state (the Lyman limit) and $\alpha_{Q}$ is the spectral index, $\Gamma^{tot} = \Gamma^{QSO}(L_{Q}, \alpha_{Q}; r) + \Gamma^{bkg}(z)$, where
\begin{align}
\label{gamma qso}
\Gamma^{QSO}(L_{Q}, \alpha_{Q}; r) 
& 
\approx
\int_{\nu_{Ly}}^{\infty}
d\nu
\sigma(\nu)
\frac{L_{\nu}}
{4 \pi r^{2} h_{P} \nu}
\\ &
=
\frac{A_{0} L_{Q}}
{4 \pi h_{P} (\alpha_{Q} + 3) r^{2}}
.
\end{align}
Here, $h_{P}$ is Planck's constant and the hydrogen photoionization cross section is approximately given by
\begin{equation}
\label{photoionization cross section}
\sigma(\nu) = A_{0} 
\left(
\frac{\nu}
{\nu_{0}}
\right)^{-3}
\end{equation}
for $\nu$ near $\nu_{Ly}$, where \mbox{$A_{0}=6.30 \times 10^{-18}$ cm$^{2}$} \citep{2006agna.book.....O}.
In equation (\ref{gamma qso}), we assume that the proper distance to the quasar, $r$, is sufficiently small that the point is effectively at the same redshift as the quasar, $z \approx z_{Q}$.

The $N-$body dark matter simulation provides the position and velocity of
each particle in the box.
These are interpolated on a uniform grid using a cloud-in-cell (CIC)
algorithm \citep{1988csup.book.....H} to construct density and
velocity fields.
Triangular-shaped cloud (TSC) interpolation is then used to smoothly
project the fields onto sight lines which are drawn through the box with randomly selected orientations.
We obtain fields $\delta(r)$ and $v_{\parallel}(r)$ ($v_{\parallel}>0$ indicates motion toward the observer) for the density and
velocity along the line of sight, respectively, as a function of proper distance from the quasar.
From the the density field, we compute the corresponding temperature field using the equation of state (\ref{igm eos}).
Given quasar properties $Q=\{z_{Q}, L_{Q}, \alpha_{Q}\}$, $\Gamma^{QSO}$ is computed at each pixel using equation \ref{gamma qso}. For simplicity, in \S \ref{proximity effect biases} -- \S \ref{halo mass} 
all simulated quasars are given the
median luminosity from the \cite{2000ApJS..130...67S} sample (see \S \ref{conventions}).

We assume that the baryons trace the dark matter.\footnote{See e.g., \cite{2001MNRAS.324..141M} and \cite{2006MNRAS.367.1655V}, for
comparisons between Ly$\alpha$ forest statistics extracted from dark matter, pseudo-hydrodynamic, and fully
hydrodynamic simulations.}
Although no Jeans-scale smoothing filter is explicitly applied, our simulation has resolution comparable to the Jeans scale ($\sim$40 km s$^{-1}$ and $\sim$20 km s$^{-1}$, respectively, at $z=3$) and we do model thermal broadening in our mock spectra (see \S \ref{redshift space distortions}).
The HI number density along the line of sight is then given by equation (\ref{hi number density}).
Ignoring redshift-space distortions for the moment and for a negligible line width, the \Lya~optical depth at any given point is
\begin{equation}
\label{tau GP}
\tau =
\frac{\pi e^{2} f_{Ly \alpha}}
{m_{e} H_{0} \nu_{Ly \alpha}}
\frac{n_{\textrm{HI}}}
{\sqrt{\Omega_{m}(1 + z)^{3} + \Omega_{\Lambda}}},
\end{equation}
where $f_{Ly \alpha}$ is the oscillator strength and $\nu_{Ly\alpha}$ is the frequency of the \Lya~transition.
Here, the optical depth as a function of distance from the
quasar, $\tau(r)$ (equivalently, the transmission coefficient, $e^{-\tau}$), is the quantity of interest and this is what we
will refer to as a ``spectrum.''

\begin{figure}[ht]
\begin{center}
\includegraphics[width=0.475\textwidth]{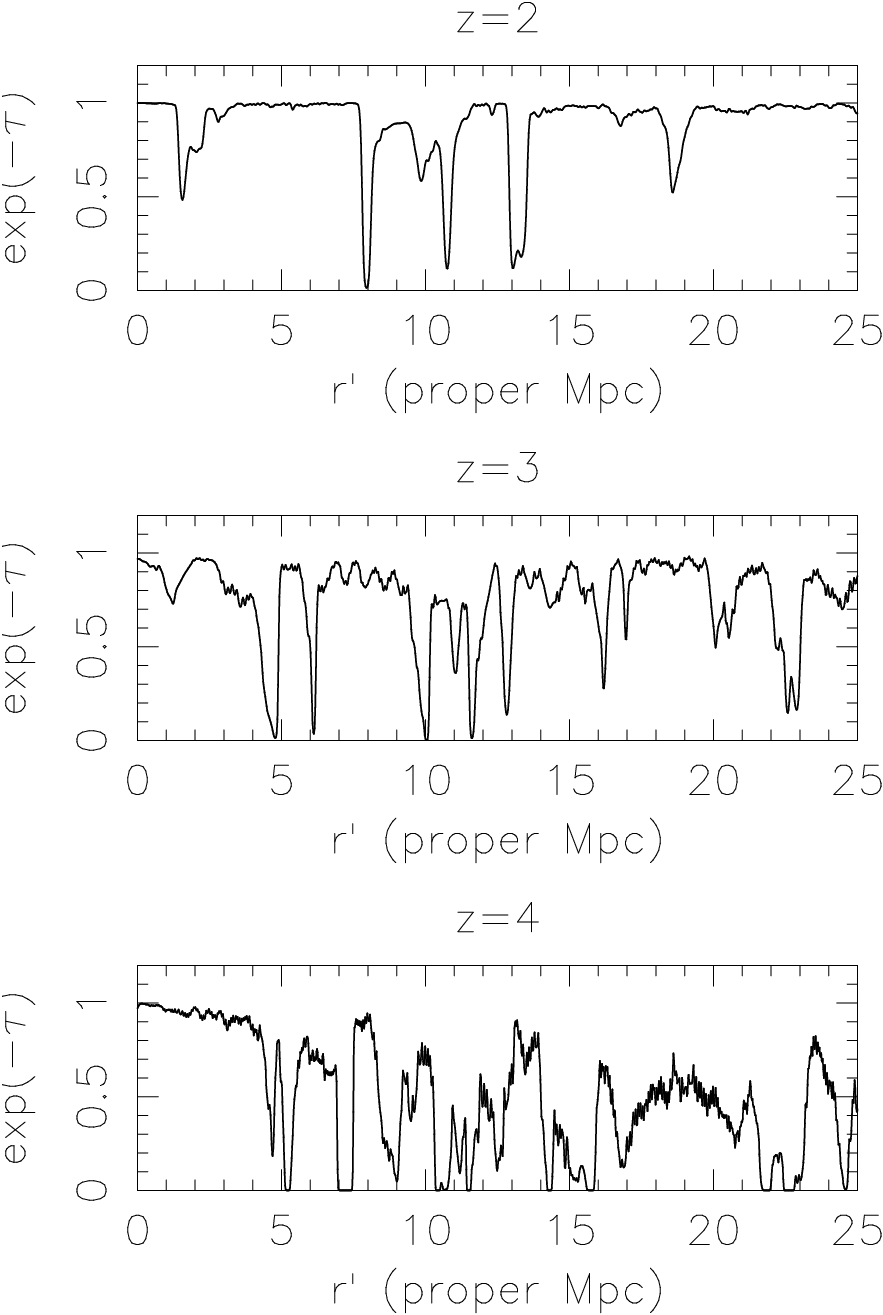}
\end{center}
\caption{Examples of mock spectra (shown as the transmission coefficient as a function of the redshift-space distance from the quasar) constructed from the simulation at at $z=2,~3,~\textrm{and}~4$ (top, middle, and bottom panel).
In each case, the quasar has typical luminosity and \mbox{$\Gamma^{bkg}=10^{-12}~\textrm{s$^{-1}$}$}.
The local overdensities in which the quasars reside and the
redshift-space distortions are modeled as in \S
\ref{overdensities} and \S \ref{redshift space distortions}.
}
\label{sample spectra}
\end{figure}

\begin{figure}[ht]
\begin{center}
\includegraphics[width=0.475\textwidth]{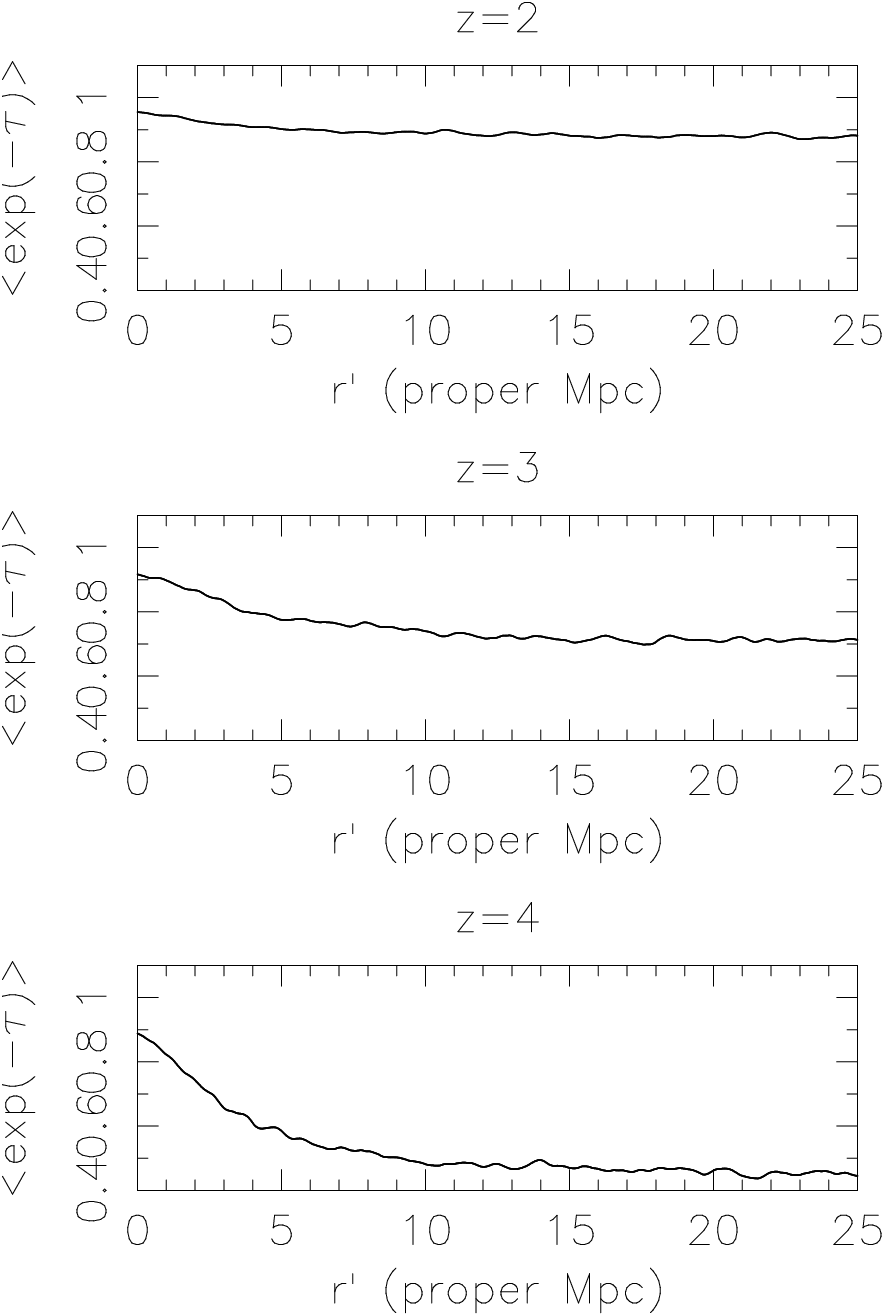}
\end{center}
\caption{Transmission coefficient as a function of redshift-space
distance from the quasar averaged over samples of 1000 mock spectra
for quasars of typical luminosity and
\mbox{$\Gamma^{bkg}=10^{-12}~\textrm{s$^{-1}$}$} at $z=2,~3,~\textrm{and}~4$ (top, middle, and bottom panel).
The proximity effect, seen as an increase in the mean transmission
near the quasars, is stronger at higher redshifts where the absorbing
gas is denser owing to the cosmological expansion.
The local overdensities in which the quasars reside and the
redshift-space distortions are modeled as in \S
\ref{overdensities} and \S \ref{redshift space distortions}.
In each case, the curves were smoothed with a boxcar of width 0.2 comoving Mpc.
}
\label{mean transmission}
\end{figure}

Figure \ref{sample spectra} shows examples of mock spectra constructed from the simulation.
Figure \ref{mean transmission} shows the transmission coefficient as a function of distance from the quasar averaged over ensembles of 1000 quasars at $z=2,~3,~\textrm{and}~4$.
The increase in the mean transmission near the quasars illustrates the
proximity effect.
We note that the effect is more pronounced at higher redshifts, even for
fixed \Gbkg~and $\Gamma^{QSO}$, owing to the increase in the neutral
hydrogen density attributed to the cosmological expansion.
At low redshifts, the IGM is more dilute and hence the transmission is
high even away from quasars.
This suggests, as corroborated by our analysis in \S \ref{llhd biased}, that it is easier to obtain constraints on \Gbkg~from
the proximity effect at higher redshifts. On the other hand, bright quasars are rarer, and continuum fitting is
more difficult, at high redshift.

\section{OVERDENSITIES, INFALL, AND CLUSTERING AROUND QUASARS}
\label{overdensities infall and clustering}
After setting up the basis of our numerical framework in the previous
section, we are now ready to discuss the effects of overdense quasar
environments on mock spectra.
We describe how we model matter overdensities in \S \ref{overdensities} and the redshift-space distortions owing to gas infall in \S \ref{redshift space distortions}.
We then pause to take a look at how the optical depth PDFs are
affected by the overdensities and redshift-space distortions
associated with massive dark matter host halos in \S
\ref{optical depth pdfs for different halo masses}.
The biases they induce on measurements of \Gbkg~are given full consideration
later in \S \ref{proximity effect biases}, after describing our
likelihood formalism.
We estimate the effect clustering of galaxies and other AGN around proximity effect quasars on measurements of \Gbkg~in \S \ref{clustering}, concluding that that it is relatively unimportant.

\subsection{Overdensities}
\label{overdensities}
Structures in the Universe such as galaxies and quasars form in the collapse of regions of the Universe with mean density exceeding the critical value necessary to overcome the Hubble flow.
Such regions preferentially arise in larger-scale overdensities \citep[e.g.,][]{1991ApJ...379..440B}.
In addition, the density enhancement around collapsed objects
generally extends well beyond their virial radius
\citep{2004MNRAS.347...59B, 2006ApJ...645.1001P}.
As a consequence, quasars are expected to be surrounded by an excess of absorbing gas with respect to a random location in the Universe.
In fact, there is strong and growing evidence from clustering measurements that
quasars reside in massive dark matter halos.
Using AGN-AGN clustering measurements from the 2dF QSO Redshift Survey (2QZ) at $0.8<z<2.1$, \cite{2004MNRAS.355.1010P} 
find a minimum mass for a DM halo to host a quasar of $10^{12}$~\Msun, with a characteristic mass $\sim 10^{13}$~\Msun.
This result has been recently confirmed by \cite{Porciani:2006tf}.

Similarly, \cite{2004ApJ...610L..85W} conclude, from a
sample of narrow-line AGN from the Sloan Digital Sky Survey (SDSS) Data Release 1, that the minimum host DM halo mass is $2 \times 10^{12}$~\Msun.
From 2QZ AGN-AGN clustering data at $0.3<z<2.2$, \cite{2005MNRAS.356..415C} find
that quasars lie in DM halos of mass $3.0 \pm 1.6~h^{-1}\times 10^{12}$~\Msun, regardless of luminosity.
\cite{2006ApJ...644..671C} measure the AGN-galaxy cross-correlation
function at $0.7<z<1.4$, using data from the SDSS and Deep
Extragalactic Evolutionary Probe (DEEP) 2 survey and find a
minimum DM halo mass $\sim5 \times 10^{11}$~\Msun, with mean $3 \times
10^{12}$~\Msun, again with no evidence for luminosity dependence.
The lack of dependence of clustering (and hence of DM halo mass) on quasar
luminosity has been confirmed at higher redshifts by \cite{2005ApJ...630...50A}, who show that the AGN-galaxy cross-correlation length at $1.8<z<3.5$ is constant over a range of 10 optical magnitudes. 
Finally, \cite{2003Natur.421..341B} estimate the DM host halo masses of the bright quasars $\textrm{SDSS}1122-0229$ at $z=4.795\pm0.004$ and $\textrm{SDSS}1030+0524$ at $z=6.28\pm0.02$ to be \mbox{$2.5\times10^{12}$~\Msun} and \mbox{$4.0\times10^{12}$~\Msun}, respectively, independently of clustering based on the spectral signature of gas infall.
\cite{2006astro.ph.11792H} compare clustering of quasars and galaxies as a function of
luminosity, redshift, and color and find that the clustering of
local ellipticals is in accord with models which associate quasar
activity with the formation of spheroids \citep{2006ApJS..163...50H}.

We model the overdensities in which quasars reside by putting each
mock quasar at the center of mass of a DM halo with mass $M_{DM}$ in
a specified range $[M_{min}, M_{max}]$ randomly chosen in the simulation box.
To identify halos in the box, we use a friends-of-friends algorithm \citep[e.g.,][]{1985ApJ...292..371D, 2003MNRAS.339..312S} with linking length $b=0.2 \times \langle \Delta r_{p} \rangle$, where the mean interparticle spacing $\langle \Delta r_{p} \rangle \equiv L / N$ for a box with side length $L$ and $N^{3}$ particles.
The mass of the halo is then the sum of the particle masses.

\begin{figure}[ht]
\begin{center}
\includegraphics[width=0.48\textwidth]{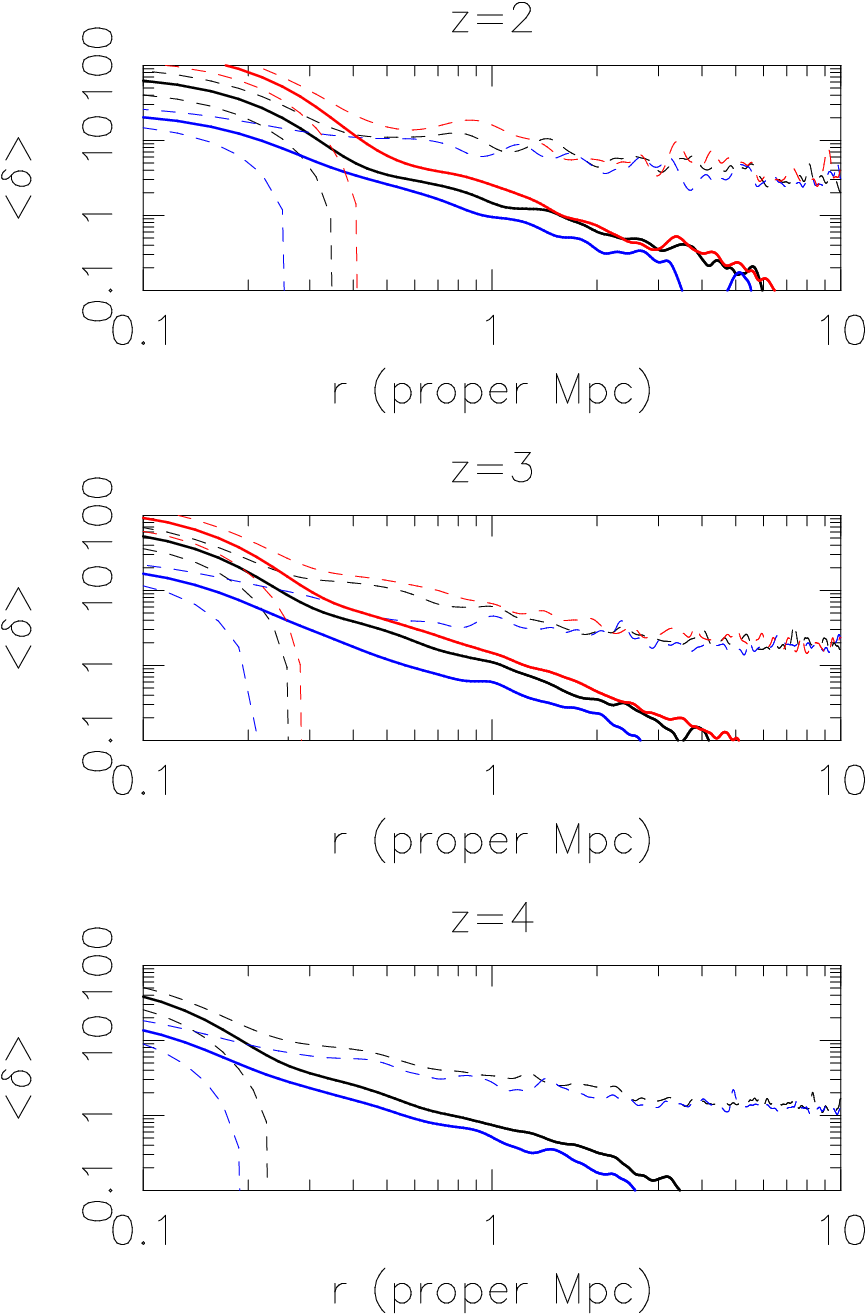}
\end{center}
\caption{Profiles of linear overdensity of dark matter halos at $z=2,~3,~\textrm{and}~4$ (top, middle, and bottom).
In each case, 100 lines of sight were drawn from each of 100 halos
with mass in the ranges \mbox{$3.0\pm1.6~h^{-1}\times10^{11}$ \Msun} (light, blue), \mbox{$3.0\pm1.6~h^{-1}\times10^{12}$ \Msun} (Croom et al. 2005, black), and \mbox{$6.0\pm3.2~h^{-1}\times10^{12}$ \Msun} (heavy, red) randomly selected in the simulation box.
The solid curves show the profiles averaged over all lines of
sight and the bounding dashed curves show the sample standard
deviations of the density at each point.
The profiles extend significantly above the mean cosmic density ($\delta \equiv 0$) at distances greatly exceeding the halo virial radii ($\sim50$ proper kpc).
We do not show the case of the most massive halos at $z=4$ because
the simulation box contains too few halos in this mass range.
}
\label{density profiles}
\end{figure}

Using this definition, \cite{2001MNRAS.321..372J} obtained a universal
mass function.
Assuming that halos are isothermal spheres, this corresponds to a mean
density inside the halo of 180 times the background density, although
in practice there is a large scatter about this value \citep{2002ApJS..143..241W}.
For our \mbox{$100~h^{-1}$ comoving Mpc} box with $256^{3}$ particles,
\mbox{$b \approx 0.08~h^{-1}$ comoving Mpc}.
We require that each halo contains at least 31 particles.
Figure \ref{density profiles} shows the mean overdensity profiles of dark
matter halos hosting quasars for the fiducial mass range \mbox{$3.0\pm1.6~h^{-1}\times 10^{12}$ \Msun}~inferred from clustering measurements by \cite{2005MNRAS.356..415C}, with the corresponding standard deviation at each point.
Also shown are analogous curves for some of the 
least (fiducial mass range divided by ten, \mbox{$3.0\pm1.6~h^{-1}\times10^{11}$ \Msun}) and most massive
halos (fiducial mass range multiplied by two, \mbox{$6.0\pm3.2~h^{-1}\times10^{12}$ \Msun}) in the simulation box.
For comparison, the virial radius of a halo is given by\footnote{This
expression is valid at $z\gtrsim1$, where the effect of a cosmological
constant can be neglected.}
\begin{equation}
\label{virial radius}
r_{vir} \approx
42 \left( \frac{M_{DM}}{10^{12}~h^{-1} \textrm{M}_{\odot}} \right)^{1/3}
\left( \frac{1 + z}{4} \right)^{-1}
\textrm{kpc}
\end{equation}
\citep{2001PhR...349..125B}, which in all cases is much smaller than the radius up to which the halo profiles have a significant impact on the density field, given that most of the \Lya~forest arises from fluctuations of order unity.
As we show the next section, the halo
infall regions extend to even larger distances.

\subsection{Infall and Redshift-Space Distortions}
\label{redshift space distortions}
\begin{figure}[ht]
\begin{center}
\includegraphics[width=0.48\textwidth]{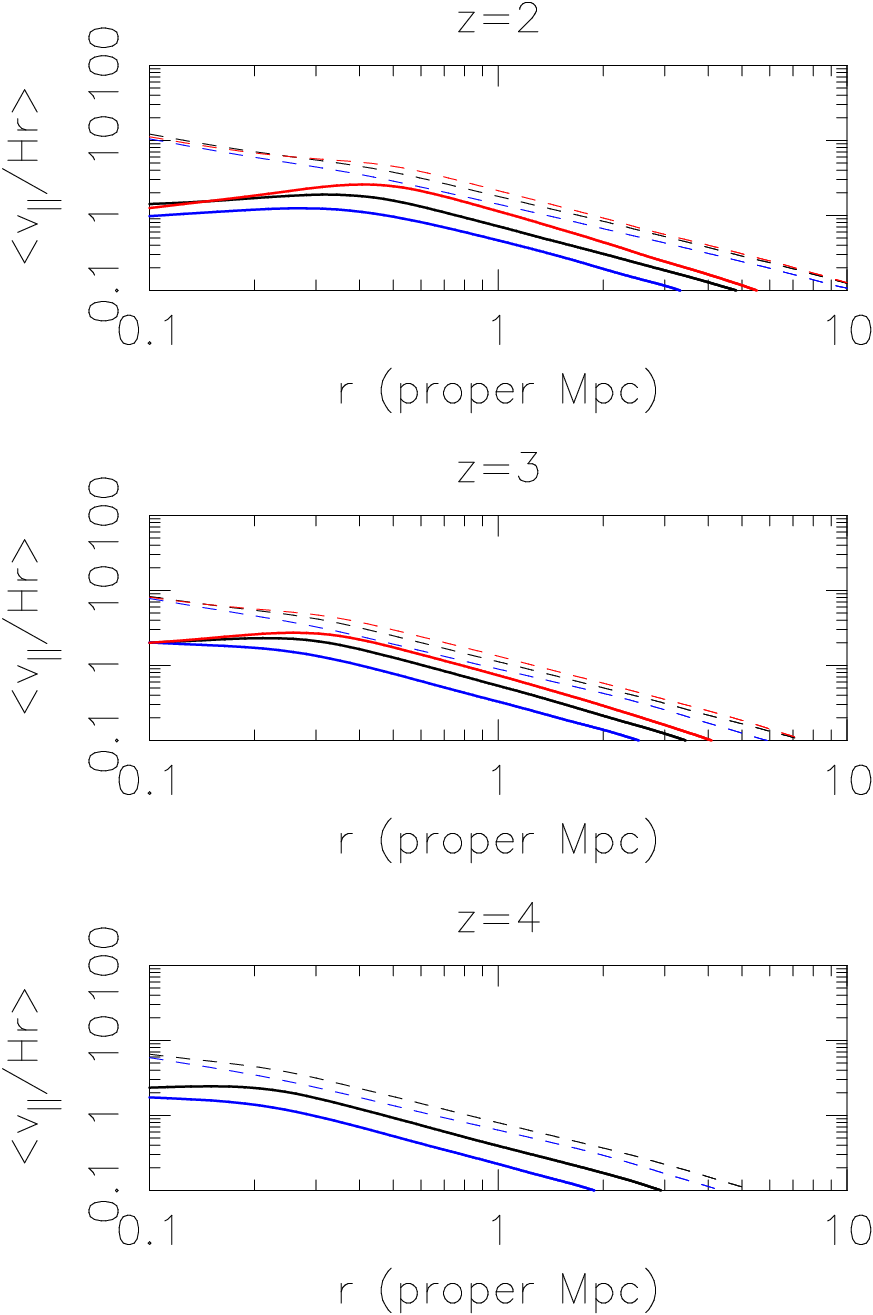}
\end{center}
\caption{Profiles of relative velocity along the line of sight between the gas and the quasar for halos at $z=2,~3,~\textrm{and}~4$ (top, middle, and bottom).
In each case, 100 lines of sight were drawn from each of 100 halos
with mass in the ranges \mbox{$3.0\pm1.6~h^{-1}\times10^{11}$ \Msun} (light, blue), \mbox{$3.0\pm1.6~h^{-1}\times10^{12}$ \Msun} (Croom et al. 2005, black), and \mbox{$6.0\pm3.2~h^{-1}\times10^{12}$ \Msun} (heavy, red) randomly selected in the simulation box.
The solid curves show the profiles averaged over all lines of
sight and the bounding dashed curves show the sample standard
deviation of the velocity at each point.
Shown are the absolute values of the true mean velocities, which are negative owing to systematic infall of the gas toward the halo centers.
The velocities are given in units of $Hr$ because in linear theory
$v_{\parallel}/Hr \sim \delta$ (c.f. Figure \ref{density profiles}) and this is the quantity with which the expected \Gbkg~bias in proximity effect analyses scales (\S \ref{relation between biases}).
We do not show the case of the most massive halos at $z=4$ because
the simulation box contains too few halos in this mass range.
}
\label{velocity profiles}
\end{figure}

In general, neither the quasar nor the absorbing gas is at rest with
respect to the Hubble flow.
In particular, quasar host halos grow via the infall of matter toward
their centers.
This is illustrated in Figure \ref{velocity profiles}, in which we show profiles of relative velocity between the quasar and the absorbing gas computed from the simulations.
The peculiar velocities have the consequence of shifting the effective
redshift of \Lya~absorption of gas parcels through the Doppler effect.

Consider the case where the quasar follows the Hubble flow and a gas parcel
along the line of sight at actual proper distance $r$ from the quasar has peculiar velocity $v_{\parallel}$ ($>0$ for motion toward the observer) along the line of sight.
This case is sufficiently general for our purpose, since the peculiar
motion of the quasar is simply taken into account by considering its
redshift as determined from spectroscopy.
The important quantity is the relative velocity of the gas with
respect to the quasar along the line of sight; in practice, the velocity of each mock quasar is identified with the velocity of the center of mass of the halo in which it lies.

To first order, the gas parcel will absorb gas that has \Lya~frequency (in the quasar rest frame) at proper distance $r ' = r + \Delta
r$ from the quasar, where $\Delta r = v_{\parallel}/H$.
We call $r'$ the ``redshift-space'' distance.
Inserting reasonable numbers ($z_{Q}=3$, \mbox{$v_{\parallel}$=300 km s$^{-1}$}; see Figure \ref{velocity profiles}), we obtain $\Delta r \approx 1$ proper Mpc.
The Doppler shift effect is thus potentially very significant.
We incorporate the effect of peculiar velocities in the usual manner, which we review here briefly
for completeness (see e.g., \citealt{1997ApJ...486..599H} for more details).
Since the ionized fraction of a given gas parcel depends on the photoionization rate at its actual proper distance $r$ from the quasar, we first calculate the neutral hydrogen density field ignoring redshift-space distortions as in \S \ref{mock spectra}.
Because of discretization, we obtain a sequence of neutral densities $\{ \rho_{0}, \rho_{1}, ..., \rho_{N} \}$, with corresponding peculiar velocities $\{ v_{\parallel,0}, v_{\parallel,1}, ..., v_{\parallel,N} \}$ at proper distances $\{ r_{0}, r_{1}, ..., r_{N} \}$ from the quasar.
For each $i$, we calculate the redshift-space distance
$r'_{i}=r_{i}+\Delta r_{i}$, with $\Delta r_{i}=v_{\parallel,i}/H$,
for the gas parcel.
The redshift-space density field $\{\rho'_{0}, \rho'_{1}, ...,
\rho'_{N} \}$ is then constructed such that 
\begin{equation}
\rho'_{i} = 
\sum_{r'_{j}>0 \textrm{ is nearest to }r_{i}} \rho_{j}
\end{equation}
i.e. at each point $i$ of the discretized line-of-sight field we sum the densities of
the gas parcels whose redshift-space coordinate is positive (the
absorption occurs along the line of sight to the quasar) and nearest
to $r_{i}$.

The thermal motion of the gas particles also causes redshift-space
distortions, broadening the absorption lines owing to the random velocities.
The characteristic temperature of the IGM is $T_{IGM} \approx 2\times10^{4}$ K
\citep{2000ApJ...534...41R, 2000MNRAS.318..817S, 2001ApJ...562...52M}, corresponding to typical Doppler thermal velocities
\mbox{$v_{rms} \approx \sqrt{3 k_{B} T_{IGM} / m_{p}} \approx 20$ km s$^{-1}$}.
This is an order of magnitude less than the typical bulk
velocities of the gas parcels owing to peculiar velocities, suggesting
that this effect has little direct impact on \Gbkg measurements from the proximity effect.
However, it is important to simulate thermal broadening to accurately
reproduce the statistical properties of the \Lya~forest.

The Doppler profile owing to thermal broadening is given by
\begin{equation}
\phi(\nu) = 
\frac{1}
{\sqrt{\pi} \Delta \nu_{D}}
\exp{
\left[
-\frac{(\nu - \nu_{\textrm{Ly}\alpha})^{2}}
{(\Delta \nu_{D})^{2}}
\right]
},
\end{equation}
with
\begin{equation}
\Delta \nu_{D}=
\frac{\nu_{\textrm{Ly}\alpha}}{c}
\sqrt{
\frac{2 k_{B} T}
{m_{p}}
}
\end{equation}
for \Lya~absorption by hydrogen atoms \citep[e.g.,][]{1979rpa..book.....R}.
A shift $\Delta \nu$ in frequency corresponds to a shift \mbox{$\Delta r = c
\Delta \nu / H \nu$} in effective proper position of absorption, so
that in proper real-space the profile becomes
\begin{equation}
\phi(\Delta r) = 
\frac{H c \nu_{\textrm{Ly}\alpha}}
{\sqrt{\pi} \Delta \nu_{D}}
\exp{
\left[
-\frac{(H c \nu_{\textrm{Ly}\alpha} \Delta r)^{2}}
{(\Delta \nu_{D})^{2}}
\right]
}.
\end{equation}
For each gas parcel we use the temperature $T$
as given by the equation of state (\ref{igm eos}) for the undistorted
density field.
For each $j$, we then assign a fraction \mbox{$\phi(r_{j} - r_{i})
\Delta r_{i}$} of the neutral density of parcel $i$ to the cell at
position $r_{j}$, where $\Delta r_{i}$ is the proper spacing between neighboring cells.
When modeling both peculiar velocities and thermal broadening
simultaneously, the numbers of cells by which each partial gas parcel
is shifted owing to peculiar motion and thermal broadening are added.
Note that we assume a thermal profile here, and hence ignore the natural
line width, but this is accurate for the low column density Ly-$\alpha$ forest
(e.g., \citealt{1997ApJ...486..599H}).

\subsection{Optical Depth PDF vs. Halo Mass}
\label{optical depth pdfs for different halo masses}
\begin{figure*}[th]
\begin{center}
\includegraphics[width=0.95\textwidth]{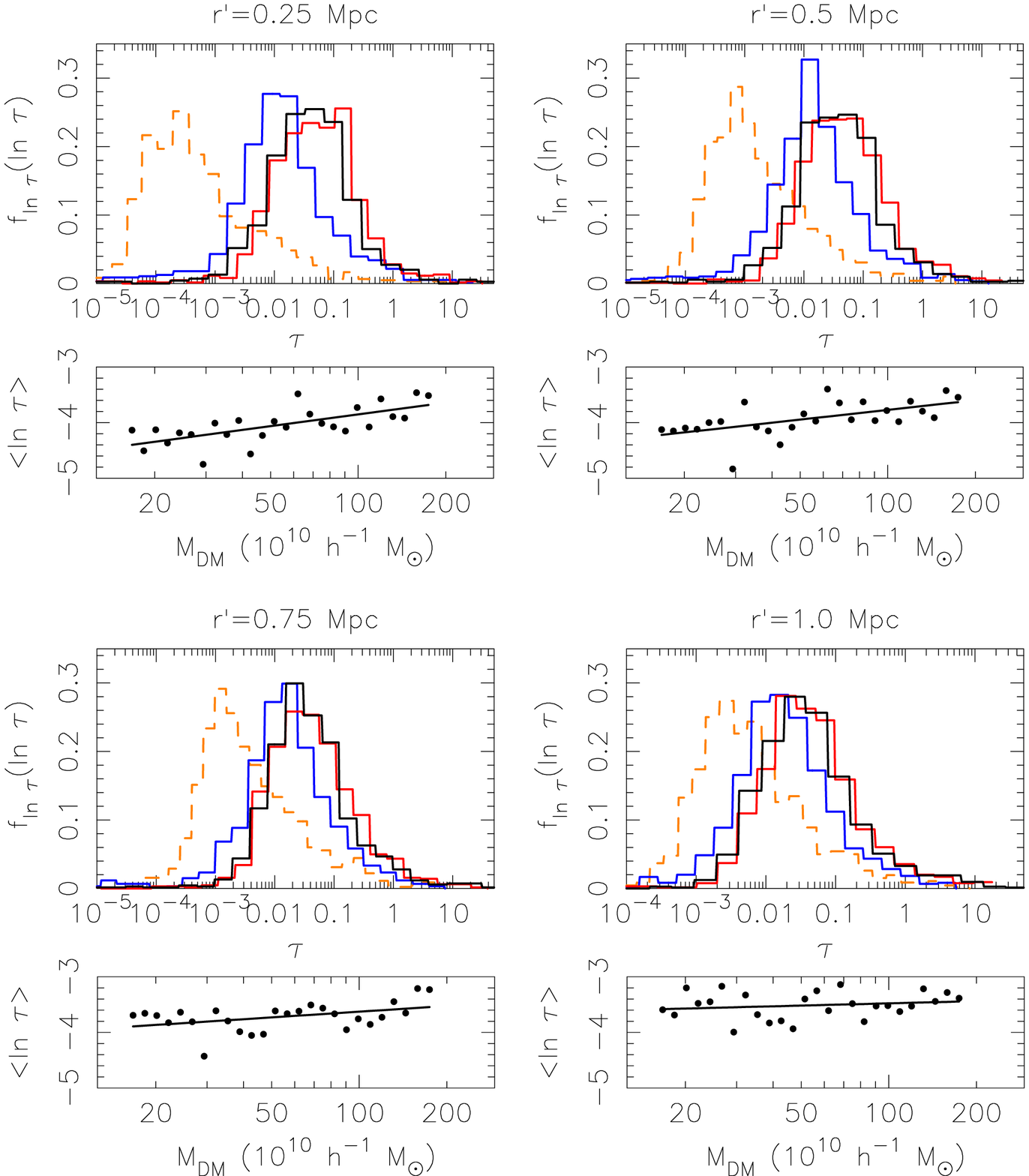}
\end{center}
\caption{
Optical depth PDFs at increasing redshift-space proper
distance from quasars of typical luminosity at $z=3$.
In each case, the solid-line histograms show PDFs approximated from 10 lines of sight drawn from each of 100 halos
with masses in the ranges \mbox{$3.0\pm1.6~h^{-1}\times10^{11}$ \Msun} (light, blue), 
\mbox{$3.0\pm1.6~h^{-1}\times10^{12}$ \Msun} (Croom et al. 2005, black), 
and \mbox{$6.0\pm3.2~h^{-1}\times10^{12}$ \Msun} (heavy, red) randomly selected in the simulation box.
The dashed orange PDFs show the case of quasars lying in random locations in the box.
This situation can be clearly distinguished from the cases were the quasars lie in massive dark matter halos.
Note, however, that although the optical depth statistics near quasars differ by large \emph{factors} for different host halo masses, the absolute optical depth differences are small, of order $\Delta \tau \sim 0.01$.
We discuss in \S \ref{practical considerations} observational difficulties which make such differences challenging to measure in practice.
In general, the optical depth PDFs are shifted to higher values as the halo
mass is increased.
Below each panel, we show how the sample mean $\ln{\tau}$ varies with mass for 10 lines of sight from each of 10 halos in bins of logarithmic width $\Delta \ln{M_{DM}}=0.05$ (10 times finer than the fiducial Croom et al. mass range) along with the least-squares log-log linear fit.
The slope of the linear fit compared to the scatter between the dots
provides an estimate of how well halos of different masses
can be distinguished with 100 lines of sight, using only one data point
at the given distance from each spectrum.
The results of full mock mass likelihood analyses are presented in \S
\ref{halo mass}.
}
\label{optical depth pdfs}
\end{figure*}
At this point, it is interesting to take a look at how the optical depth
PDF varies with halo mass at different distances from the quasars
when the overdensity and gas infall associated with dark matter host
halos are modeled as above.
This is illustrated in Figure \ref{optical depth pdfs}, in which we
show results for dark matter halos of mass \mbox{$3.0\pm1.6~h^{-1}\times10^{11}$ \Msun} (light), \mbox{$3.0\pm1.6~h^{-1}\times10^{12}$ \Msun} (Croom et al. 2005), and \mbox{$6.0\pm3.2~h^{-1}\times10^{12}$ \Msun} (heavy) at $z=3$, along with fits to the trend of mean $\ln{\tau}$ with halo mass for the range of masses probed by our simulation.
We also show the case in which quasars lie in random locations, which
can be clearly distinguished from the massive halos.
The optical depth PDFs are shifted to higher values with increasing halo mass, which is expected since the matter overdensity should increase with halo mass.
The significant effect of quasar environments on the optical depth
statistics in the proximity regions of quasars suggests that existing proximity
effect analyses which have neglected these will not necessarily
recover the correct \Gbkg.
We make analytic estimates of the expected biases in \S \ref{analytic bias estimates} and quantify these more accurately with mock spectra in \S \ref{llhd biased}.
The environmental dependence of the optical depth statistics also raises the possibility of using the latter to probe quasar environments, in particular measure the mass of the host halos.
We focus attention to this exciting possibility in \S \ref{halo
mass}.
For the moment, we note that the effect of halos on the optical depth
statistics is to ``separate'' the PDFs with mass.
This is nondegenerate with varying \Gbkg, which just shifts PDFs of
$\ln{\tau}$ by the same amount, regardless of the halo mass\footnote{Ignoring redshift-space distortions, $\tau \propto n_{\textrm{HI}} \propto {(\Gamma^{tot}})^{-1}$ (\S \ref{mock spectra}).}.
We must note that although the optical depth statistics near quasars differ by large \emph{factors} for different host halo masses, the absolute optical depth differences are small, of order $\Delta \tau \sim 0.01$.
Measuring these differences in actual spectra, with noise and for which the continuum level must be estimated from the data, may be challenging in practice, as we discuss in \S \ref{practical considerations}.

\subsection{Clustering}
\label{clustering}
A possible bias
related to overdensities is the clustering of galaxies and other quasars around proximity effect quasars.
If quasars indeed form in large-scale overdensities, then the probability of other quasars and galaxies forming in the vicinity is increased with respect to average regions of the Universe, because there it is easier for local density peaks (``peaks within peaks'') to cross the threshold for collapse \citep{1991ApJ...379..440B}.
The clustering of galaxies around quasars has in fact been both predicted in simulations of galaxy and quasar formation \citep[e.g.,][]{2002MNRAS.332..529K} and observationally measured over a wide redshift interval (e.g., Croom et al. 2004\nocite{2004ASPC..311..457C} at $z<0.3$; Coil et al. 2006\nocite{2006ApJ...644..671C} at $z\sim1$; Adelberger \& Steidel 2005a\nocite{2005ApJ...630...50A} at $1.5 \lesssim z \lesssim 3.5$). 
Quasars, which are generally thought to be hosted by galaxies, have
been shown to exhibit similar clustering among themselves
\citep[e.g.,][]{2001MNRAS.325..483C, 2004ASPC..311..457C, 2004ApJ...610L..85W, 2005MNRAS.356..415C}. 

We have thus far identified \Gbkg~with the total photoionization rate
near quasars contributed by all \emph{other} light sources, implicitly
assuming that this quantity is equal to the total photoionization rate
averaged over large regions of the Universe (the ``true'' background
rate).
Since galaxies and quasars (which together presumably dominate the contribution to the photoionizing background) cluster around quasars, we in fact expect \Gbkg~near
quasars to be higher than the true background \Gbkg.
In this section, we quantitatively estimate this effect.

Let us first suppose that the ionizing background is dominated by emission from star-forming LBGs, as may be the case at $z\sim3$ \citep[e.g.,][]{2001ApJ...546..665S}.
Consider the photoionizing flux at a given position owing to all sources of light in the Universe \emph{other} than the quasar in an idealized model of $N_{G}$ isotropic sources, each with
specific luminosity $L_{\nu,G}$, distributed in a spherical volume
$V=4 \pi R^{3} / 3$ with quasar-galaxy correlation function 
\begin{equation}
\label{quasar galaxy correlation function}
\zeta_{QG}(r) = 
\left(
\frac{r}
{r_{0}}
\right)^{-\gamma}.
\end{equation}
Cosmological expansion can be neglected in the argument, since for
$\gtrsim2$ the radiation is largely local, i.e. sources at higher
redshifts are absorbed and can be neglected \citep{1999ApJ...514..648M}.
Let $n_{G}$ be the mean number density of sources and
$f_{\nu,G}=L_{\nu,G} / 4 \pi r^{2}$ be the specific flux at distance $r$ from each source.
Suppose that a quasar is located at the origin of the volume.
The total specific flux owing to the other sources at a
point displaced by the vector $\bs{r}$ from the quasar is
\begin{equation}
f_{\nu,G}^{other} =
\sum_{i=1}^{N_{G}} f_{\nu}^{i} =
\frac{L_{\nu,G}}
{4 \pi}
\sum_{i=1}^{N_{G}} |\bs{r - r_{i}}|^{-2},
\end{equation}
where the sums are over the sources other than the quasar and $\bs{r_{i}}$ is the position of source $i$.
Since we will eventually use this quantity for the sole purpose of
calculating a photoionization rate, we need not worry that light rays
incident at $\bs{r}$ are not necessarily parallel, i.e. the scalar sum of the
specific fluxes contains all the relevant information.
To first order, we have, averaging over $\bs{r_{i}}$,
\begin{equation}
\langle f_{\nu,G}^{other} \rangle (r) =
\frac{N_{G} L_{\nu,G}}
{4 \pi}
\langle |\bs{r - r_{i}}|^{-2} \rangle,
\end{equation}
where
\begin{align}
\label{mean d2}
\langle |\bs{r - r_{i}}|^{-2} \rangle 
&
=
V^{-1}
\int_{0}^{R}
\int_{0}^{\pi}
\int_{0}^{2\pi}
d(|\bs{r}-\bs{r_{i}}|)
d\theta
d\phi
\sin{\theta}
\\ &\times
[1 + \zeta_{QG}(r_{i})]
.
\end{align}
The integral above is over a sphere of radius $R$ centered on $\bs{r}$: the $|\bs{r}-\bs{r_{i}}|^{2}$ factor of the spherical volume element cancels the $|\bs{r}-\bs{r_{i}}|^{-2}$ term that is integrated over.
Using $r_{i}^{2}=r^{2} + |\bs{r - r_{i}}|^{2} - 2 r |\bs{r - r_{i}}| \cos{\theta}$ (for a suitable choice for the orientation of the coordinate system),
we may evaluate
\begin{equation}
\label{mean f_nu other integral}
\langle f_{\nu,G}^{other} \rangle (r) =
\langle f_{\nu,G}^{other,smooth} \rangle +
\langle f_{\nu,G}^{other,clust} \rangle(r),
\end{equation}
where
\begin{equation}
\langle f_{\nu,G}^{other,smooth} \rangle=n_{G} L_{\nu,G} R
\end{equation}
is the contribution of the smooth background component (arising from
the factor 1 in eq. \ref{mean d2}) and
\begin{equation}
\langle f_{\nu,G}^{other,clust} \rangle = 
\frac{n_{G} L_{\nu,G}}
{2}
\int_{0}^{R}
\int_{0}^{\pi}
d(|\bs{r} - \bs{r_{i}}|)
d\theta
\sin{\theta}
\zeta_{QG}(r_{i})
\end{equation}
is the component (arising from the $\zeta_{QG}$ term in eq. \ref{mean d2}) owing to the clustering of galaxies around the quasar.
For both the smooth and clustering components, we define the
photoionization rates as calculated from the average contributions:
\begin{equation}
\Gamma^{bkg,x} \equiv
\int_{\nu_{Ly}}^{\infty}
d\nu \sigma(\nu) 
\frac{
\langle f_{\nu,G}^{other,x} \rangle
}
{h_{P} \nu}.
\end{equation}

Are the local fluctuations in the photoionizing flux owing to the
quasar-galaxy correlations important when attempting to measure
\Gbkg~using the proximity effect?
To answer this question, we note that the proximity effect is
sensitive to $\Gamma^{bkg}=\Gamma^{bkg,smooth}+\Gamma^{bkg,clust}$ and consider the ratio:
\begin{equation}
\label{clustering to smooth ratio}
\frac{\Gamma^{bkg,clust}}
{\Gamma^{bkg,smooth}}(R; r)
=
\frac{\int_{0}^{R}
\int_{0}^{\pi}
d(|\bs{r} - \bs{r_{i}}|)
d\theta
\sin{\theta}
\zeta_{QG}(r_{i})
}
{2 R}
.
\end{equation}
Note that this ratio is purely geometrical, depending on only the
attenuation radius and the quasar-galaxy correlation function.
In particular, it is independent of the abundance and brightness of
the galaxies.
The cancellations occur because we are assuming that the smooth
background and the fluctuations are produced by the same sources.

\begin{figure}[ht]
\begin{center}
\includegraphics[width=0.575\textwidth,angle=-90]{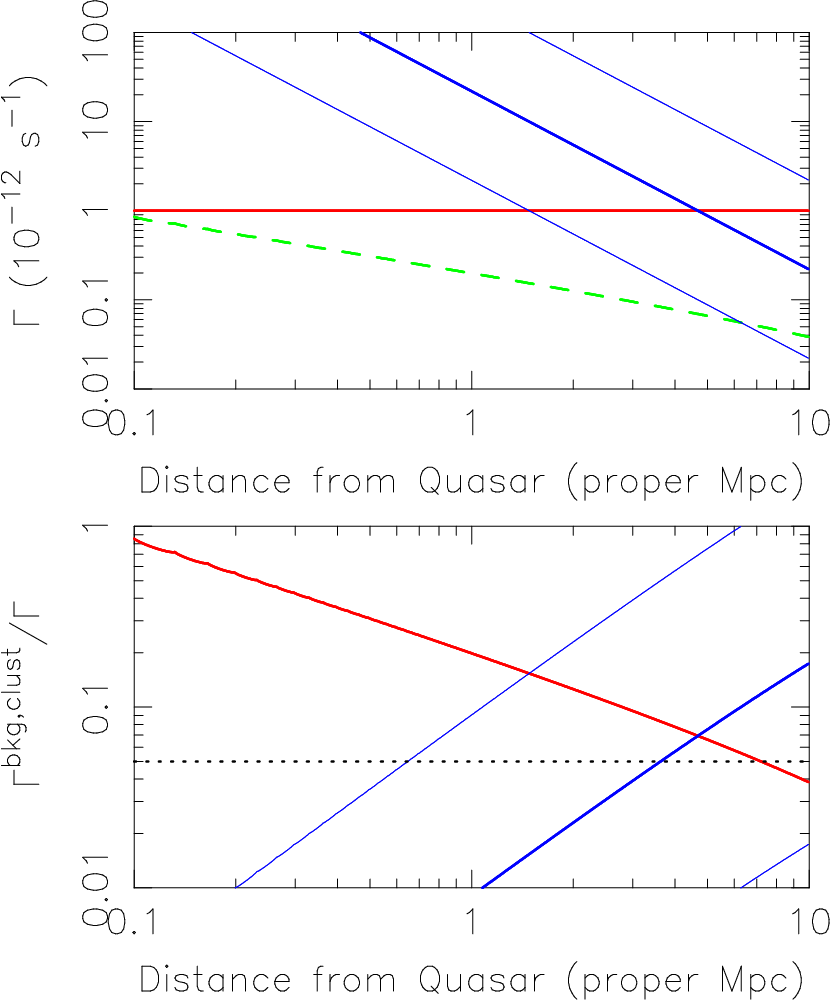}
\end{center}
\caption{
Top: Comparison of the contributions to the total hydrogen photoionizing
rate owing to the background (\mbox{$\Gamma^{bkg}=10^{-12}$
s$^{-1}$}; horizontal red line), the quasar (solid blue curves; the
central thick curve corresponds to the mean typical quasar luminosity
and the bounding thin curves indicates ranges of one standard
deviation), and the local excess of galaxies owing to AGN-galaxy
clustering (assuming that the background is dominated by emission from LBGs; green dashed curve) as a function of distance from the quasar at $z=3$.
Bottom: Ratios of the local contributions from the quasar (blue) and the
clustering of galaxies around the quasar (red) to the true photoionizing
background.
The horizontal dotted line shows the \mbox{5\%} level.
}
\label{Gamma comparison}
\end{figure}

In the top panel, Figure \ref{Gamma comparison} shows $\Gamma^{bkg,smooth}$, $\Gamma^{bkg,fluct}$, and $\Gamma^{QSO}$ as a function of distance from the quasar for the order-of-magnitude estimate \mbox{$\Gamma^{bkg}=10^{-12}$~s$^{-1}$} (see Figure \ref{existing measurements figure}) and quasars of typical luminosities at $z=3$.
For the quasar-galaxy correlation function, we adopt \mbox{$r_{0}=5$ h$^{-1}$}
comoving Mpc and $\gamma=1.6$, consistent with the results of \cite{2005ApJ...627L...1A} for LBGs at $z \sim 3$.
The characteristic value of the $R$ is the mean free path $l_{mfp}$ of ionizing
photons, i.e. the radius beyond which incoming ionizing photons are
exponentially suppressed.
\cite{1999ApJ...514..648M} calculate 
\mbox{$l_{mfp} \approx 33 [(1 + z)/4]^{-4.5}$} proper Mpc for Lyman limit photons, which we take as characteristic of ionizing soft galactic emission.

In the bottom panel, the ratios of $\Gamma^{bkg,fluct}$ to $\Gamma^{bkg,smooth}$ and $\Gamma^{QSO}$ are shown.
It is seen that $\Gamma^{bkg,fluct}\ll\Gamma^{bkg,smooth}$ ($<20$\% for \mbox{$r>1$ Mpc} and $<5$\% for \mbox{$r>7$ Mpc}, which is much less than the factor $\gtrsim2$ discrepancy between proximity effect and flux decrement measurements of \Gbkg), except at very small distances from the quasar, where it diverges. 
This divergence is most likely benign, as it occurs on a scale much
smaller than the proximity region (i.e., the region where $\Gamma^{bkg} \sim \Gamma^{QSO}$, which contains significant information about $\Gamma^{bkg}$) and in any case is
an artifact of assuming that the correlation function (eq. \ref{quasar
galaxy correlation function}) maintains its power-law behavior as
$r\to0$.
In reality, $\Gamma^{bkg,fluct}$ is expected to be bounded above on
physical grounds. Moreover, note that in the region where $\Gamma^{bkg,fluct} \sim \Gamma^{bkg,smooth}$, the 
quasar flux already dominates over the locally-enhanced background flux by a large factor.
We thus conclude that clustering of galaxies around quasars is
unlikely to significantly affect measurements of \Gbkg~using the
proximity effect, at least at $z=3$ for a background which is dominated by the contribution from LBGs.

What if quasars contribute significantly to \Gbkg~and how is the
situation modified at different redshifts?
Consider the case of a background dominated by quasars.
Then we must consider the quasar-quasar correlation function, but it
is in fact similar to the quasar-LGB one
\citep[e.g.,][]{2005MNRAS.356..415C}.
Let us denote it generically by $\zeta(r)$.
The ratio in equation (\ref{clustering to smooth ratio}) is then only
modified by the attenuation length $R$. For a galaxy-dominated background, we took
$R$ to be the mean-free path to Lyman-limit photons, but the appropriate $R$ for
quasars will be larger owing to their harder spectra and the longer mean free paths
for high energy quasar photons.
Since $\zeta(r)$ is a strictly decreasing function of $r$,
\begin{equation}
\left(
\frac{\Gamma^{bkg,clust}}
{\Gamma^{bkg,smooth}}
\right)_{QQ}
<
\left(
\frac{\Gamma^{bkg,clust}}
{\Gamma^{bkg,smooth}}
\right)_{QG},
\end{equation}
where the subscripts $QQ$ and $QG$ refer to the quasar- and
galaxy-dominated cases, respectively.
The clustering effect is therefore even less important in the
quasar-dominated case.
For the general case in which both galaxies and quasars contribute
significantly to the background, \Gbkg~is a linear superposition of the
galaxy and quasar contributions and so the total clustering
contribution is again negligible. 
For redshifts $z<3$, the attenuation length is increased because
the number density of absorbers is decreased by cosmic expansion, so the result still holds.
In the limit of large redshifts, $R\to0$ and
$\Gamma^{bkg,clust}/\Gamma^{bkg,smooth} \to \infty$, so that the
result must eventually break down.
However, we are here mainly concerned with explaining the
discrepancy between proximity effect and flux decrement measurements
of $\Gamma^{bkg}$ at $z\lesssim3$ (Figure \ref{existing measurements figure}).
The argument given so far indicates that clustering has a negligible
effect and we are therefore not compelled to model it and to pursue a
further analysis of the issue.

\section{BIASES IN PROXIMITY EFFECT MEASUREMENTS OF \Gbkg}
\label{proximity effect biases}
To understand why the proximity effect
measurements of \Gbkg~are systematically higher than the flux
decrement measurements of the same quantity, we first formulate a statistical method for estimating \Gbkg~from the proximity effect, with physical assumptions mimicking those made in previous proximity effect measurements of \Gbkg.
In contrast to most previous analyses, we consider the optical depth
statistics in the proximity region of quasar spectra instead of the
differential number of absorption lines as a function of redshift,
$dN/dz$.
Following the seminal work of \cite{1988ApJ...327..570B}, the differential number of absorption lines has been used in almost all studies of the proximity effect to date.
However, absorption lines arising from discrete absorbers are no longer
consistent with the modern picture of the \Lya~forest as arising from
smooth density fluctuations in the IGM.
Moreover, the counting of absorption lines is a somewhat ill-defined
and uncertain procedure, making it difficult to rigorously treat
statistically.
On the other hand, once the quasar continuum is estimated, the optical depth at any point can be simply measured as \mbox{$\tau=-\ln{F^{obs}/F^{cont}}$} \footnote{Because of the redshift-space distortions, this quantity is not directly proportional to HI density as in equation (\ref{tau GP}). It is best interpreted as a redshift-space effective optical depth, or taken as the definition of the optical depth in this work.}, where $F^{obs}$ is the observed flux and $F^{cont}$ is the continuum level.
As we will see, it is also easy to treat the optical depth in a
statistically sound framework.

To estimate the biases introduced by the local overdensities around and the
infall of gas toward the quasars (the two effects which are most
likely to play an important role according to our discussion in
\S \ref{mock spectra}) in analyses which have neglected those
complications, 
we apply our measurement method to mock spectra (with a known value of \Gbkg) in which those effects are modeled.
We begin with some remarks on the optical depth probability
density function (PDF) in the general \Lya~forest in \S
\ref{optical depth distribution function} and formulate our statistical
formalism in \S \ref{Gbkg ML}.
In \S \ref{analytic bias estimates}, we provide
order-of-magnitude analytic estimates of the expected biases, before
quantifying those more accurately using mock spectra in \S
\ref{llhd biased}. 

\subsection{Lyman$-\alpha$ Forest Optical Depth PDF}
\label{optical depth distribution function}
Before proceeding with defining a method for inferring \Gbkg~from
optical depth statistics in the proximity region, it is useful to consider the optical depth PDF in the general \Lya~forest.
\cite{1991MNRAS.248....1C} first used a lognormal PDF to describe the
dark matter density field and showed that it follows plausibly from the assumption of initial linear Gaussian and velocity fluctuations.
\cite{1992A&A...266....1B} and \cite{1997ApJ...479..523B} applied this approach to baryons also and showed that the resulting \Lya~forest matched the observations well.
Several observational studies, e.g. \cite{2006astro.ph..7633B} and \cite{2007MNRAS.374..206D}, have also found that the optical depth PDF in the \Lya~forest is close to lognormal. 

\begin{figure}[ht]
\begin{center}
\includegraphics[width=0.475\textwidth]{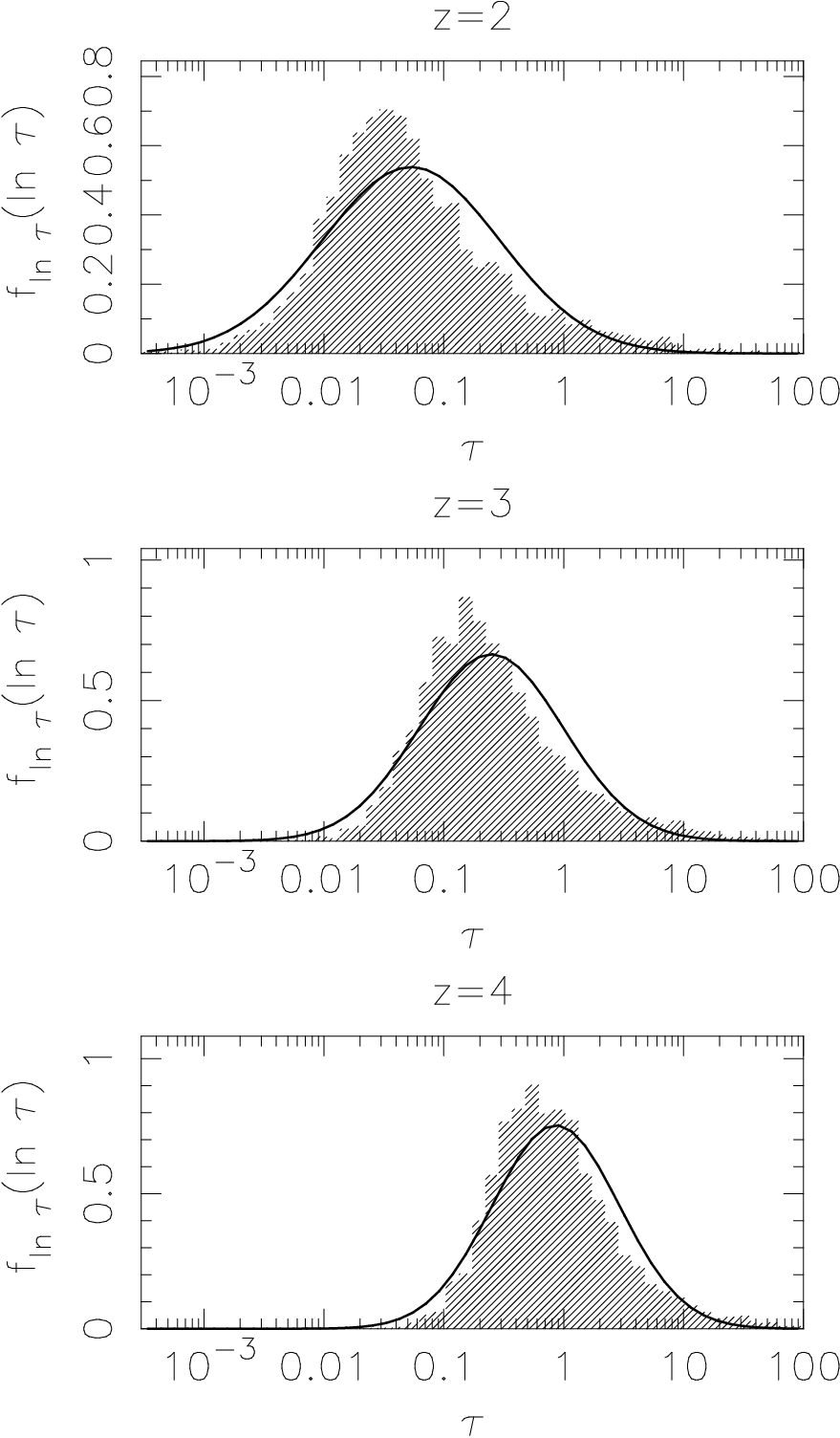}
\end{center}
\caption{Histograms of the optical depth as constructed from random
lines of sight in the simulation box at $z=2,~3,~\textrm{and}~4$ (top,
middle, and bottom panels).
Redshift-spaces distortions have been modeled as in \S
\ref{redshift space distortions}.
The solid curves show the corresponding lognormal
distribution as estimated from the mean logarithm of the optical depth
and its standard deviation.
}
\label{tau hists}
\end{figure}

Figure \ref{tau hists} shows normalized histograms of the optical
depth constructed from random lines of sight through our simulation
box at $z=2,~3,~\textrm{and}~4$, along with the corresponding lognormal
distribution as estimated from the mean logarithm of the optical depth
and its standard deviation.
Although there are noticeable deviations -- especially at low redshifts -
of the simulated optical depth distribution from lognormal shape, the
main features of the distributions are well-captured by a
lognormal.
The lognormal provides a convenient analytic form, significantly simplifying the discussion, and so we will use it in making estimates of the biases that arise when naively measuring \Gbkg~from the proximity effect.
Our results should be robust to this assumption as, in this analysis,
the information about \Gbkg~is contained in the \emph{scaling} equation
(\ref{simple tau scaling}).
\subsection{Maximum Likelihood Method for \Gbkg}

\label{Gbkg ML}
We now formulate a maximum likelihood method to measure \Gbkg~from
optical depth statistics in the proximity regions of Ly$\alpha$ absorption spectra.
While the statistical formalism differs from the BDO line-counting method, we make essentially the same physical assumptions.
Thus, our conclusions should also provide insight in the validity of studies
which used the line-counting approach, i.e. virtually all existing proximity effect
analyses.
Mathematically, the explicit assumption is that that the \Lya~optical depth in the presence of the quasar, $\tau^{prox}$, is related to
the optical depth if the quasar were turned off, $\tau^{off}$, by
\begin{equation}
\label{simplest photoionization eq}
\tau^{prox}
=
\frac{\tau^{off}}
{1 + \omega(Q; r)},
\end{equation}
where $\omega(Q; r)=\Gamma^{QSO}(Q; r)/\Gamma^{bkg}(z_{Q})$, 
$\Gamma^{QSO}(Q; r)$ is the contribution to the total photoionization rate of the quasar at distance $r$ from it, and 
$Q=\left\{ z_{Q}, L_{Q}, \alpha_{Q} \right\}$ specifies the redshift, luminosity and spectral index of the quasar under consideration.
We call this method the ``$\tau-$scaling''
likelihood analysis.
Owing to the small size of the proximity region, $r$ can be taken to be the proper distance.
Equation (\ref{simplest photoionization eq}) is just what would be
obtained from photoionization equilibrium assuming that the quasar
lies in a random location and neglecting all redshift-space distortions.
In what follows, we simplify the notation and use $\tau$ to denote $\tau^{off}$.

Using equation (\ref{simplest photoionization eq}), we can relate the
optical depth statistics in the \Lya~forest away from the quasar, but at
approximately equal redshift ($z \approx z_{Q}$), to those inside the proximity region.
For a lognormal $\tau$ distribution with mean logarithm $\langle \ln{\tau}  \rangle$ and standard deviation, also in the logarithm, $\sigma_{\ln{\tau}}$, we have:
\begin{equation}
f_{\widehat{\tau}}(z_{Q}; \tau)=
\frac{1}
{\sqrt{2 \pi} \sigma_{\ln{\tau}} \tau}
\exp{
\left[ 
-
\frac{(\langle \ln{\tau}  \rangle - \ln{\tau})^{2}}
{2 \sigma_{\ln{\tau}}^{2}}
\right]
}
,
\end{equation}
which translates to lognormal distribution in the
proximity region given by:
\begin{align}
\label{tau prox distribution}
&
f_{\widehat{\tau}^{prox}}(Q, r; \tau^{prox})=
\frac{1 + \omega(Q; r)}
{\sqrt{2 \pi} \sigma_{\ln{\tau}} \tau^{prox}} 
\notag \\
& \times 
\exp{ 
\left[ 
-
\frac{(\langle \ln{\tau}  \rangle - \ln[{(1 + \omega(Q; r))\tau^{prox}}])^{2}}
{2 \sigma_{\ln{\tau}}^{2}}
\right]
}
.
\end{align}
We define a correlation length, $r_{corr}$, such that points separated
by this distance can be assumed to have independent optical depths.
Then, for a given quasar spectrum, we estimate $\langle \ln{\tau} \rangle$ and $\sigma_{\ln{\tau}}$ using the usual unbiased estimators
\begin{equation}
\label{mean tau}
\widehat{ \langle \ln{\tau} \rangle } =
\frac{1}{N}
\sum_{i=1}^{N} \ln{\tau_{i}}
\end{equation}
and
\begin{equation}
\label{sigma tau}
\widehat{ \sigma_{\ln{\tau}} }^{2} =
\frac{1}{N-1}
\sum_{i=1}^{N} \left(\ln{\tau_{i}} - \widehat{ \langle \ln{\tau} \rangle } \right)^{2},
\end{equation}
where $\tau_{i}$ are optical depths at points separated by $r_{corr}$ outside the proximity region.

We may now construct the likelihood function
\begin{equation}
\label{simplest likelihood}
\mathcal{L}[\Gbkg(z)]
\equiv
\prod_{Q}
\left\{
\prod_{prox~pts}
f_{\widehat{\tau}^{prox}}(\tau^{prox})
\right\},
\end{equation}
where $f_{\widehat{\tau}^{prox}}(\tau^{prox})$ has the same parameter
dependence as in equation (\ref{tau prox distribution}).
Here, the outer product is over the different quasars in a sample.
The inner product is over data points in the proximity region of quasar $Q$ (defined to be between $r_{min}$ and $r_{max}$ from the quasar), again separated by a proper distance $r_{corr}$.
Each factor is proportional to the probability of obtaining a realization of the optical depth given the adopted model for \Gbkg.
The terms in the products are assumed to be evaluated at independent points,
so that the likelihood is proportional to the probability of obtaining
all the measured optical depths given the model.
By Bayes's theorem, this is proportional to the probability that the
assumed model is correct.
After normalization, the likelihood function gives the PDF for a given model for \Gbkg~to be correct.
The likelihood estimator in equation \ref{simplest likelihood} is suboptimal, since it does not make use of correlated data points.
The estimated likelihood may therefore be wider (less constraining) than could be achieved in principle.
The design of an optimal estimator is outside of the scope of this paper. 
In \S \ref{practical considerations}, we discuss practical considerations arising with actual quasar spectra of finite resolution and signal-to-noise ratio, which are likely to be more important than the optimity of the likelihood estimator.

\subsection{Analytic Estimates of Biases}
\label{analytic bias estimates}
Before applying the $\tau-$scaling likelihood to mock
spectra, we make order-of-magnitude estimates for the biases that may
be expected when the method is applied to spectra with quasars lying
in overdense regions and with redshift-space distortions.
Our estimates are based on the fact that the likelihood analysis is
sensitive to the $(1 + \omega(Q; r))^{-1}$ scaling assumed for $\tau$
in the proximity region (eq. \ref{simplest photoionization eq}).
The local overdensities and infall regions around quasars distort
this scaling, leading to an incorrect estimate of $\omega$.
Assuming that $\Gamma^{QSO}(r)$ is known (e.g., from measurements of the
magnitude and spectrum of the quasar), this leads to an incorrect
estimate of \Gbkg.

\subsubsection{Bias Owing to Local Overdensity}
Consider first the case where the quasar lies in an overdense region,
but where redshift-space distortions are ignored.
Let us denote by $\omega^{true}(r)$ the true ratio \mbox{$\Gamma^{QSO}(Q; r)/\Gamma^{bkg}$} and by $\omega^{app}(r)$ the ratio as inferred when assuming the scaling given by equation (\ref{simplest photoionization eq}).
Denote the overdensity \mbox{$\Delta(r) \equiv \rho(r) / \langle \rho \rangle$}.
Then, using $\tau \propto n_{\textrm{HI}} \propto \Delta^{2 - 0.7\beta}$ (eq. \ref{hi number density}, \ref{H recombination coefficient}, and \ref{igm eos}), we expect the optical depth scaling to be approximately modified
to
\begin{equation}
\tau^{prox} = 
\frac{\Delta^{2-0.7\beta} \tau}
{1 + \omega}.
\end{equation}
Assuming the scaling 
of equation (\ref{simplest photoionization eq}) amounts to incorrectly assuming that $\Delta \equiv 1$, from which we obtain the relation
\begin{equation}
\Delta^{2 - 0.7\beta} (1 + \omega^{true})^{-1} = (1 + \omega^{app})^{-1},
\end{equation}
and hence
\begin{equation}
\frac{\Gamma^{bkg,true}}
{\Gamma^{bkg,app}}
=
\frac{\omega^{app}}
{\omega^{true}} = 
\Delta^{0.7\beta-2} (1 + {\omega^{true}}^{-1}) - {\omega^{true}}^{-1}.
\end{equation}
The typical distance $r_{typ}$ from the quasar at which information about
\Gbkg~is encoded in the proximity effect is such that the photoionizing rate owing to the quasar is comparable to the background.
We take it to satisfy $\omega^{true}(r_{typ})=1$, so that
\begin{equation}
\label{Gbkg true over app overdensity}
\frac{\Gamma^{bkg,true}}
{\Gamma^{bkg,app}} =
2 \Delta^{0.7\beta-2} - 1.
\end{equation}
It will be useful to express the bias in terms of the linear
overdensity $\delta=\Delta-1$; taking the reciprocal of equation (\ref{Gbkg true over app overdensity}), we have the bias owing to the local overdensity to first order in $\delta$:
\begin{equation}
\label{bias overdensity}
B^{od} \equiv
\frac{\Gamma^{bkg,app}}
{\Gamma^{bkg,true}}
\approx 1 + 3.1 \delta(r_{typ})
\end{equation}
for $\beta=0.62$, the late reionization limit.

\subsubsection{Bias Owing to Gas Infall}
Consider now the bias owing to the redshift-space distortions caused by
gas infall toward the centers of halos.
To simplify the estimate, we ignore the local overdensity here.
As explained in \S \ref{redshift space distortions}, a gas parcel at proper distance $r$ from the quasar and with velocity $v_{\parallel}$ along the line of sight will, owing to the Doppler effect, appear to be absorbing from a distance $r'=r + \Delta r$ from the quasar, with $\Delta r = v_{\parallel}/H$.
In this case, assuming the scaling of equation (\ref{simplest photoionization
eq}) results in the incorrect identification 
\begin{equation}
\omega^{true}(r) = \omega^{app}(r'),
\end{equation}
which gives, using $\Gamma^{QSO}(r) \propto r^{-2}$, the bias owing to
gas infall
\begin{equation}
\label{bias infall}
B^{infall}
\equiv
\frac{\Gamma^{bkg,app}}
{\Gamma^{bkg,true}}
= 
\left(
\frac{r}
{r'}
\right)^{2}
\approx
1 - 2\frac{v_{\parallel}(r_{typ})}{H r_{typ}}
.
\end{equation}

\subsection{Relation Between Overdensity and Infall Biases}
\label{relation between biases}
The density and velocity fields are related by mass conversion through
the continuity equation.
We thus expect $B^{od}$ and $B^{infall}$ to also be simply related.
In this section, we establish this connection.
For $\delta \ll 1$, we may use the linear theory continuity equation
\begin{equation}
\label{continuity}
\frac{\partial \delta}
{\partial t}
+
\nabla \cdot \bs{v}
\approx
\frac{\partial \delta}
{\partial t}
+
\frac{\partial v_{\parallel}}
{\partial r}
=
0,
\end{equation}
where $t$ is proper time, and the spatial derivatives are proper-coordinate derivatives. 
For the second equality, we have assumed pure radial infall toward the
center of the halo and neglected spherical-geometry corrections to
the divergence which are important only at small radii. 
Within the spherical collapse model \citep{1972ApJ...176....1G} and for given initial conditions, one
could write an exact relation between $\delta$ and $v_{\parallel}$
from equation (\ref{continuity}).
However, a simple expression for this relation is not available.
We thus simply proceed with an order-of-magnitude estimate, expressing
derivatives in terms of characteristic scales at $r_{typ}$:
\begin{equation}
\label{continuity oom}
\frac{|\delta|}
{t_{ff}}
\sim
\frac{|v_{\parallel}|}
{r_{typ}},
\end{equation}
where the free-fall time for a spherically symmetric mass configuration
of interior mean density $\bar{\rho}$ is given by
\begin{equation}
\label{tfree}
t_{ff}=
\frac{1}{4}
\sqrt{
\frac{3 \pi}
{2 G \bar{\rho}}
}.
\end{equation}
In fact the interior mean density exceeds the mean cosmic density
around a halo, for otherwise it would not have collapsed.
In spite of this, we use the cosmic mean density in the denominator of equation (\ref{tfree}), since
using the full overdensity would result only in a second order correction to $|\delta|/t_{ff}$ (eq. 
\ref{continuity oom}).

Solving for $|v_{\parallel}|$ in equation (\ref{continuity oom}) and substituting in equation (\ref{bias infall}) for the bias owing to infall, we obtain
\begin{equation}
\label{bias infall in terms of delta}
B^{infall} \approx 1 + 1.57 f \delta(r_{typ}).
\end{equation}
The factor $f$, of order unity, has been introduced to allow for the
approximations made.
In deriving equation (\ref{bias infall in terms of delta}), we have made use of the Friedmann equation which expresses the Hubble parameter in terms of $G\bar{\rho}$,
\begin{equation}
H=
\sqrt{
\frac{8 \pi G \bar{\rho}}
{3}
}
\end{equation}
(again neglecting the cosmological constant, which is valid at the
redshifts of interest), and of the fact that $v_{||}<0$.
Comparing the expressions for $B^{od}$ and $B^{infall}$ (eq. \ref{bias overdensity} and \ref{bias infall in terms of delta}) in terms of $\delta(r_{typ})$, we see that both biases scale similarly with $\delta(r_{typ})$.
The total bias can be estimated at $B^{tot} \approx B^{od} \times B^{infall}
\approx 1 + 4.67\delta(r_{typ})$ (for $f=1$).
At $z=3$, \mbox{$r_{typ}=5$ proper Mpc} for a quasar of typical
luminosity (Figure \ref{Gamma comparison}) and $\langle
\delta(r_{typ}) \rangle = 0.1$, with a large point-to-point dispersion of
$\sigma_{\delta} = 2$.
One may thus reasonably choose $\delta=0.32$, obtaining
$B^{tot}\approx2.5$, and quantitatively make sense of the results of
the detailed calculations of the next section.  

\subsection{Biased \Gbkg~Likelihoods}
\label{llhd biased}
\begin{figure}[ht]
\begin{center}
\includegraphics[width=0.475\textwidth]{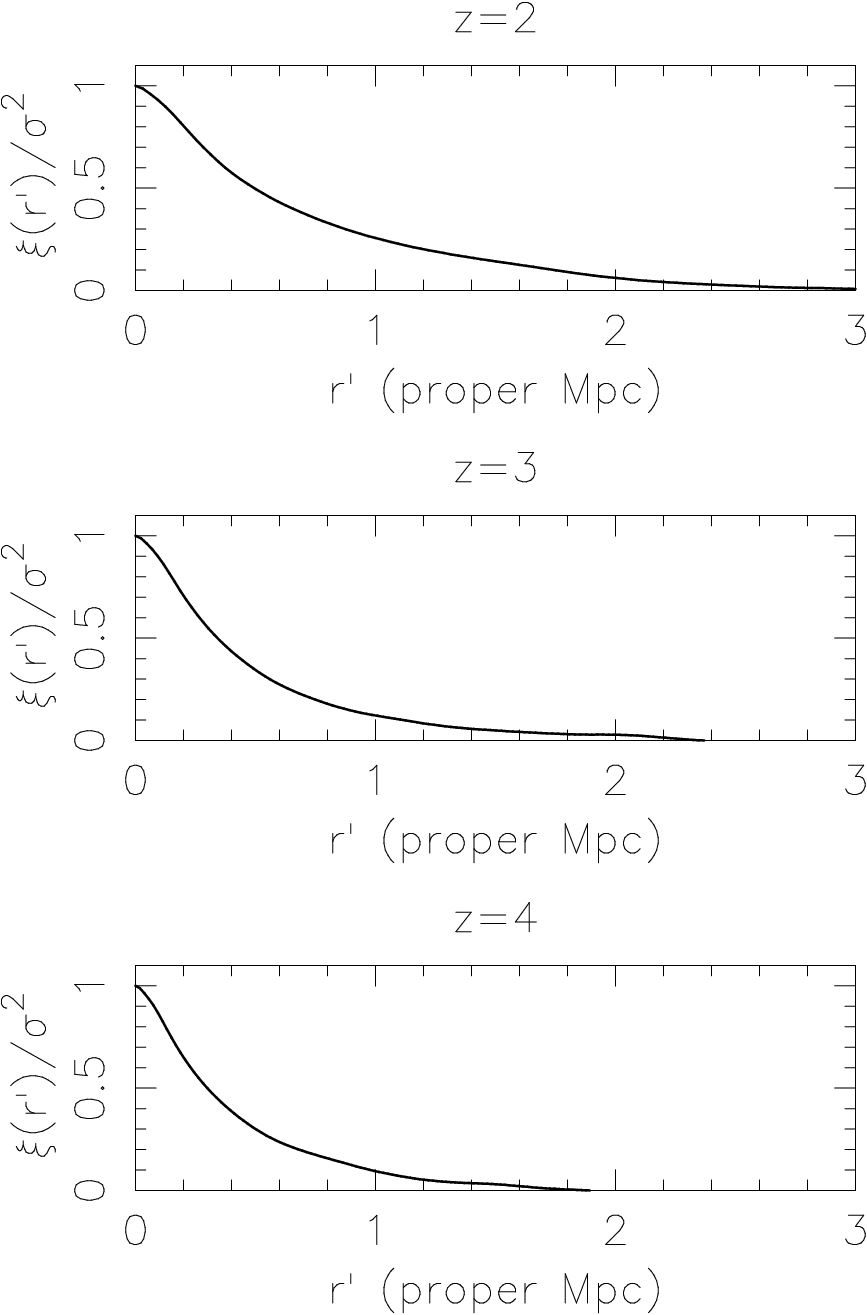}
\end{center}
\caption{
Ratio of the correlation function of $\ln{\tau}$ to its variance computed from the simulation at $z=2, 3,~\textrm{and }4$ (top, middle, and bottom).
The ratio is $<15\%$ at $z=2$ and $3$, and $<30\%$ at $z=4$ for $r'>1$ proper Mpc.
}
\label{ln tau cf}
\end{figure}

To more accurately estimate the biases that could be present in proximity effect
measurements of \Gbkg, we perform the full likelihood
analysis of \S \ref{Gbkg ML} on mock spectra with known \Gbkg~and
quasars which reside at the centers of dark matter halos with masses in
the fiducial range \mbox{$3.0\pm1.6~h^{-1}\times10^{12}$~\Msun}.
We consider both the case in which only the local overdensity is modeled
in the mock spectra and the realistic case in which the redshift-space distortions owing to gas infall and thermal broadening are also simulated.
For the likelihood analysis, we choose the parameters \mbox{$(r_{mÄin},
r_{max}, r_{corr})=(1, 20, 1)$} proper Mpc so as to include the region
where the proximity effect is important for typical quasars (e.g.,
Figure \ref{mean transmission}).
The correlation length is selected based on measurements of the
flux correlation function and power spectrum in the \Lya~forest, which indicate§ small correlations on approximately this scale \citep{2000ApJ...543....1M, 2006ApJS..163...80M}.
In Figure \ref{ln tau cf}, we show the ratio of the correlation function of $\ln{\tau}$ to its variance, $\epsilon$, computed from the simulation box (at random locations) at $z=2, 3,~\textrm{and}4$.
The ratio is $<15\%$ at $z=4$ and $3$, and $<30\%$ at $z=2$ for $r'>1$ proper Mpc.
In Appendix \ref{effect of correlations on likelihood}, we develop a toy model to estimate the effect of correlations on estimates of the likelihood that ignore them, as we calculate.
We show there that the error on the width of estimated likelihood, with respect to the true likelihood (which would take into account the correlations), is a factor $\approx(1 - \epsilon/2)$ (for $\epsilon \ll 1$, neglecting correlations between nonadjacent points), independent of the total number of (independent) spectra used.

\begin{figure}[ht]
\begin{center}
\includegraphics[width=0.475\textwidth]{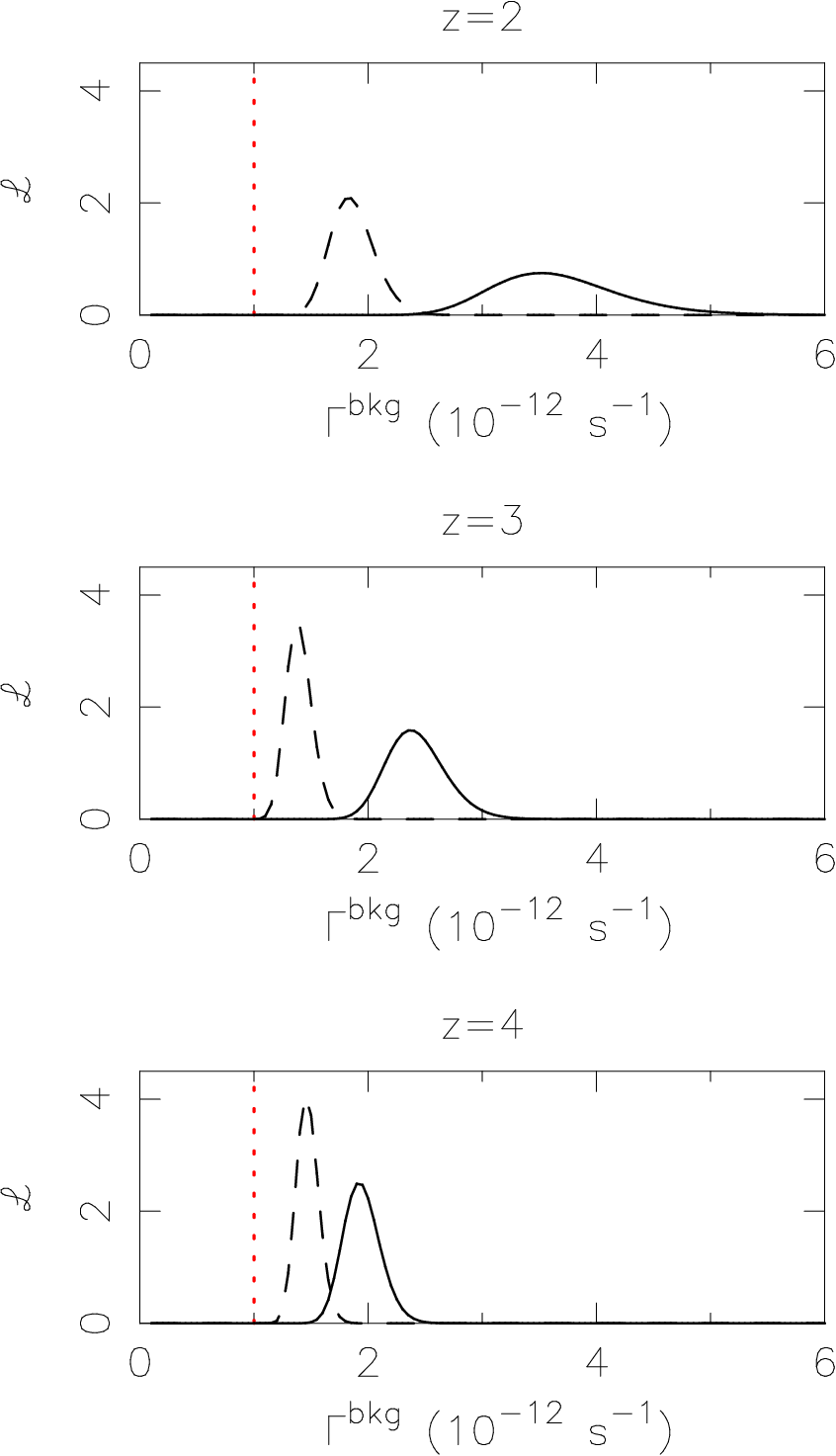}
\end{center}
\caption{
Likelihoods for \Gbkg~assuming the mapping $\tau^{prox}=\tau (1 +
\omega)^{-1}$ computed on samples of 100 mock spectra at $z=2,~3,~\textrm{and}~4$
(top, middle, and bottom) for quasars of typical luminosity lying at the centers of mass of dark matter halos of mass in the range \mbox{$3.0\pm1.6\times10^{12}~h^{-1}$~\Msun}. 
The dashed curves show the cases where the overdensities around quasars have been modeled in the mock spectra, but with no redshift-space distortions.
The solid curves show the realistic case where the mock spectra include both the overdensities and the redshift-space distortions owing to gas infall and thermal broadening.
The red vertical dotted line indicates the known \Gbkg~in the simulation
(\mbox{$10^{-12}$ s$^{-1}$}).
Deviations of the solid likelihood peaks from this value are estimates of 
the bias which is introduced in analyses, e.g. BDO-type, which ignore overdensities and redshift-space distortions.
At $z=3$, this total bias is $\approx2.5$, in quantitative agreement with the
discrepancies between existing proximity effect and flux decrement measurements
of \Gbkg~(c.f. Figure \ref{existing measurements figure}).
The overdensities and redshift-space distortions are seen to
contribute approximately equally to the total bias.
}
\label{Gbkg biased llhd}
\end{figure}

The resulting likelihood functions, computed on mock data sets of 100
typical spectra at $z=2,~3,~\textrm{and}~4$, are shown in Figure
\ref{Gbkg biased llhd}.
In each case, the maximum-likelihood \Gbkg~overestimates the true
\Gbkg, as qualitatively expected from the analytic estimates of
\S \ref{analytic bias estimates}.
Moreover, the total bias in the central ($z=3$) panel agrees quantitatively
with the analytic estimate for that redshift.
The ``typical'' $\delta(r_{typ})$ used for this estimate was in fact adjusted to quantitatively reproduce the order-of-unity bias shown here; the essential point is that the bias can be understood to order of
magnitude.
Most importantly, the above shows that the fact that quasars lie in massive host halos, associated with local matter overdensities and gas infall, \emph{alone} could bias proximity effect measurements of \Gbkg~high by a factor $\sim2.5$ at $z\approx3$.
These effects should thus be taken into account in future proximity effect analyses.

\section{UNBIASED MONTE CARLO LIKELIHOOD FOR \Gbkg}
\label{unbiased llhd}
\subsection{Method}
\label{unbiased llhd method}
The likelihood analysis in the previous section is biased because the scaling 
of equation (\ref{simplest photoionization eq}) ignores the local matter overdensity and redshift-space distortions induced by quasar host halos.
The likelihood method can however be modified to take these effects
into account and yield an unbiased measurement of \Gbkg.
To do so, it suffices to replace expression (\ref{tau prox distribution}) for $f_{\widehat{\tau}^{prox}}(\tau^{prox})$ in the likelihood function (\ref{simplest likelihood}) by PDFs numerically constructed from mock spectra with quasars placed in halos of the correct mass and with redshift-space distortions properly modeled.
The PDFs constructed in this way will match those expected in reality for the
correct value of \Gbkg.
The likelihood method will in this case be unbiased.
In this section, we demonstrate how this can be achieved.
Because the PDFs used in this method are generated by Monte Carlo, we refer to this analysis as the ``Monte Carlo'' (or MC) likelihood.

The first step is to construct optical depth PDFs from the mock spectra as a
function of redshift-space proper distance $r'$ from the quasar.
We assume here that the mass range of halos hosting quasars, $[M_{min}, M_{max}]$, is known (e.g., from clustering measurements).
For any given set of quasar parameters \mbox{$\{z_{Q}, L_{Q}, \alpha_{Q}\}$}, we generate mock spectra for quasars at redshift
$z_{Q}$ in halos in the mass range $[M_{min}, M_{max}]$, with spectral energy
distribution parameterized by $L_{Q}$ and $\alpha_{Q}$\footnote{In
fact, it is not necessary to know $L_{Q}$ and $\alpha_{Q}$ separately;
it is sufficient to give the integral quantity $\Gamma^{QSO}$ at some
distance from the quasar.}. 
We do so for each value of \Gbkg~at which we wish to evaluate the
likelihood.
In each case, 10 lines of sight are drawn from each of 100 halos randomly chosen in the simulation box.
This gives a total of 1000 lines of sight which should capture well
the variance between halos of similar masses.

Since $\tau$ is nearly lognormally distributed, it is best to first work in logarithmic space and then calculate the PDF of $\tau$ itself from the transformation
\begin{equation}
f_{\widehat{\tau}}(\tau) = 
\frac{1}{\tau}
f_{\ln{\widehat{\tau}}}(\ln{\tau}).
\end{equation}
For \mbox{$r' \in \{r_{min}, r_{min}+r_{corr}, ..., r_{max}\}$}, the
optical depths at redshift-space distance $r'$ from the quasar are
tabulated for the $10^{4}$ mock spectra.
Using the first four moments (mean, standard deviation, skewness, and
kurtosis), we obtain a very good approximate analytic expression for $f_{\ln{\widehat{\tau}}}(\ln{\tau})$ using the Edgeworth expansion, described in Appendix \ref{edgeworth expansion}.
Figure \ref{edgeworth expansion example} shows a comparison of the Edgeworth approximation to the data from which it is computed for a representative example of the PDF of the optical depth in the proximity regions of quasars of typical luminosity lying in halos in the fiducial mass range at $z=3$.
After grids of such PDFs are constructed, likelihood analysis is
performed as before, simply using these instead of the incorrectly
scaled PDFs of equation (\ref{tau prox distribution}).
The Edgeworth expansion accurately approximates the PDF of the data, as illutrated in Figure \ref{edgeworth expansion example}.
For $10^{4}$ data points at redshift-space distance $r'=0.5$ proper Mpc from quasars of typical luminosity in halos in the fiducial mass range $3.0\pm1.6\times10^{12}~h^{-1}$ \Msun~at $z=3$, the reduced $\chi^{2}$ between the data and its Edgeworth approximation is $\chi^{2}=0.87$.
In \S \ref{halo mass test}, we compute likelihoods using this approximation on mock spectra and recover the correct halo mass, thus confirming that the approximation does not introduce significant biases.
\begin{figure}[ht]
\begin{center}
\includegraphics[width=0.325\textwidth,angle=-90]{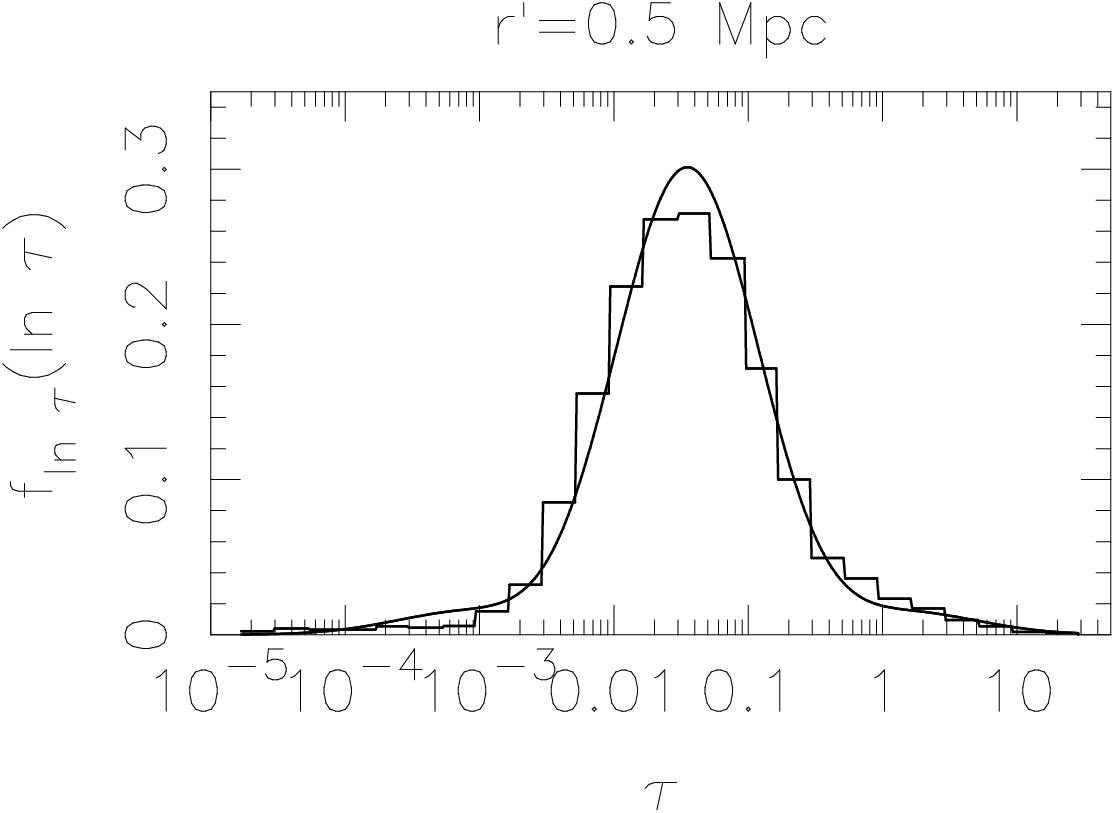}
\end{center}
\caption{
Example of the Edgeworth approximation to the optical depth PDF.
The normalized histogram shows $10^{4}$ data points at redshift-space distance $r'=0.5$ proper Mpc from quasars of typical luminosity in halos in the fiducial mass range $3.0\pm1.6\times10^{12}~h^{-1}$ \Msun~at $z=3$ in the simulation.
The Edgeworth expansion, calculated from the first four moments of the data, is shown by the solid curve.
Neglecting bins with less 500 data points (which reduce the $\chi^{2}$ even further since the tails are best approximated) and accounting for the four degrees of freedom estimated from the data, the reduced $\chi^{2}=0.87$, i.e. the Edgeworth approximation is a very good fit to the data.
}
\label{edgeworth expansion example}
\end{figure}

\subsection{Test of the Monte Carlo Likelihood}
\begin{figure}[ht]
\begin{center}
\includegraphics[width=0.475\textwidth]{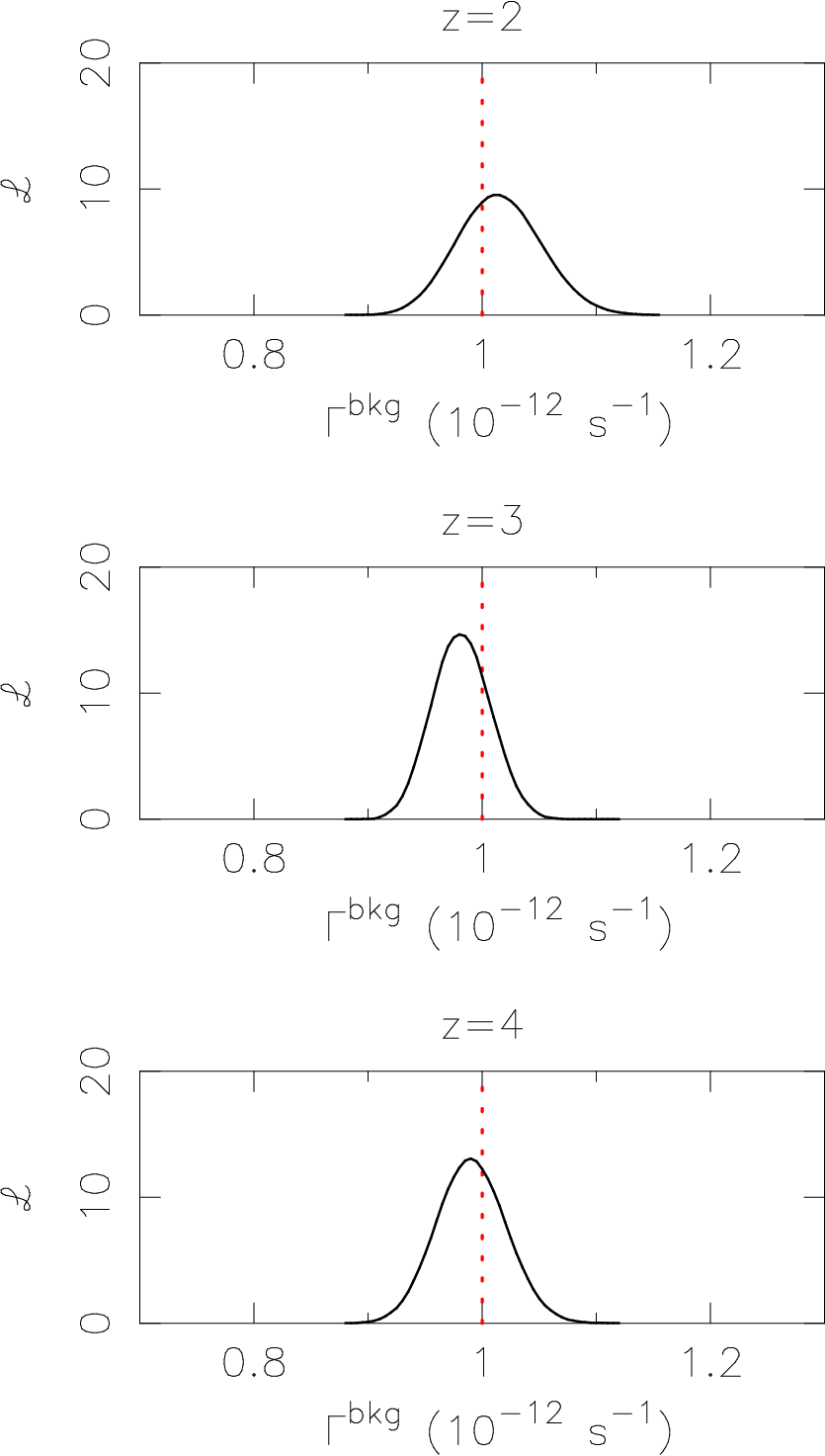}
\end{center}
\caption{
Likelihoods for \Gbkg~computed from Monte Carlo optical depth PDFs on samples of 100 mock spectra at $z=2,~3,~\textrm{and}~4$
(top, middle, and bottom) for quasars of typical luminosity lying at the centers of mass of dark matter halos of mass in the range \mbox{$3.0\pm1.6\times10^{12}~h^{-1}$~\Msun}, with redshift-space distortions and overdensities fully modeled. 
As expected, the simulated \Gbkg~(\mbox{$10^{-12}$ s$^{-1}$}, indicated by the vertical red dotted lines) is accurately recovered.
}
\label{MC llhd typical}
\end{figure}

To test the MC likelihood method, we proceed as in \S \ref{llhd
biased} for the $\tau-$scaling method: we again apply the analysis to
mock data sets with known \Gbkg~and see if the correct value is
recovered.
We test the MC likelihood on spectra with both the local
overdensity and redshift-space distortions simulated as in \S
\ref{llhd biased}.
The results are shown in Figure \ref{MC llhd typical} for 100 spectra
at $z=2,~3,~\textrm{and}~4$.
As expected, the correct value of \Gbkg~is accurately recovered.

\subsection{Relationship to Flux Decrement Method}
Unlike the $\tau-$scaling method, which is sensitive only to the
scaling of the optical depth in the proximity region with respect to
far away from the quasar\footnote{The optical depth statistics away from each quasar are estimated from the data in the analysis.}, the Monte Carlo likelihood method is
sensitive to the absolute level of absorption, both close to and far
from the quasar.
This is because the MC PDFs at each distance from the quasar depend on this absolute
level. 
In fact, in the limit where we ignore points in the proximity regions
of the quasars and extend the analysis to greater distances
($r_{prox}\ll r_{min}<r_{max}$), the method essentially reduces to the
flux decrement approach.
This proximity analysis thus, although in principle unbiased, no
longer presents clear advantages over the flux decrement method, such
as being independent of $\Omega_{b}$.

\section{MASS OF QUASAR HOST HALOS FROM THE PROXIMITY EFFECT}
\label{halo mass}
In the MC likelihood method of the previous section, we assumed that $M_{DM}$ was known and maximized the likelihood function with respect to \Gbkg.
In this section, we assume we have an independent measure of \Gbkg,
coming either from the flux decrement method or
any other reliable measurement. We then 
proceed as in the previous section, but instead parameterize the
likelihood function by the mass of quasar host halos and try to constrain it from the data. 
In what follows, we test this idea on mock spectra.

\subsection{Method}
\label{halo mass method}
PDFs are numerically constructed as in \S \ref{unbiased llhd
method},
but this time \Gbkg~is fixed and the mass range \mbox{$[M_{min}, M_{max}]$} is varied.
For each mass range, we again draw 10 lines of sight from 100 randomly
selected halos.
Examples of resulting PDFs for different redshift-space distances from
the quasars and host halo masses at $z=3$ are shown 
in Figure \ref{optical depth pdfs}.
The number of halos employed in constructing the PDFs imposes limitations on the width of the mass
ranges that can be used, since the simulation box contains a finite
number of halos in any given mass interval.
Moreover, the width of the optical depth PDF for any given mass range depends on the width of the range.
We thus consider mass ranges of fixed width in the logarithm, equal to
that of the fiducial range \mbox{$3.0\pm1.6~h^{-1}\times10^{12}~$\Msun}, covering
the range of halo masses represented in the
simulation and requiring that each mass range contains at least 100
halos.
The mass ranges are indexed by the central mass in logarithmic space, \mbox{$M_{DM}=\exp{[0.5(\ln{M_{min}} + \ln{M_{max})}}]$}.
This choice is consistent with the exponential nature of the mass
function \citep[e.g.,][]{2001MNRAS.321..372J} and avoids an excessive
pile-up of halos to the left of the mass index.

\subsection{Location of Mass Information}
\label{halo mass information}
In the case of \Gbkg, the proximity effect optical depth
statistics are expected to convey significant information in the entire region
where $\Gamma^{QSO}$ and \Gbkg~are comparable.
This region can extend to $\gtrsim20$ proper Mpc for bright quasars at
$z=3$ (Figure \ref{Gamma comparison}).
However, the halo overdensity and infall velocity (in units of $Hr$) profiles drop greatly on scales $\lesssim1$ proper Mpc (Figures \ref{density profiles} and \ref{velocity profiles}).
The information about the dark matter halos is contained in these
signatures and it is a priori unclear how far away from a quasar one
can go and still learn about its host halo.
We address this question in this section, before proceeding with
testing the likelihood method for recovering $M_{DM}$ in the next.

To quantify the mass information content as a function of distance
from a quasar, we use the Kolmogorov-Smirnov (K-S) test \citep{1992nrca.book.....P}.
This statistical test compares two random samples and quantifies the
significance that the realized values were drawn from different distributions
via the K-S P-value, $P_{\textrm{K-S}}$.
If $P_{\textrm{K-S}}<\alpha$, then a deviation as large as observed between the two random samples would have probability $<\alpha$ to occur by chance if the two samples were drawn from the same distributions.

Here, we fix $\{z_{Q}, L_{Q}, \alpha_{Q}\}$, the known host halo mass, $M_{DM}^{true}$, and the number of spectra $N_{Q}$ in the mock data set and consider points at increasing distance from the quasar.
At each redshift-space distance distance $r'$, we vary $M_{DM}$.
For each trial value, we compare the sample of $N_{Q}$ optical depths
at distance $r'$ in the mock spectra to a sample of 1000 optical
depths for $M_{DM}^{true}$.
If the resulting $P_{\textrm{K-S}}$ is small, then the optical depth
statistics can significantly distinguish between $M_{DM}$ and
$M_{DM}^{true}$.

\begin{figure*}[p]
\begin{center}
\includegraphics[width=0.7\textwidth]{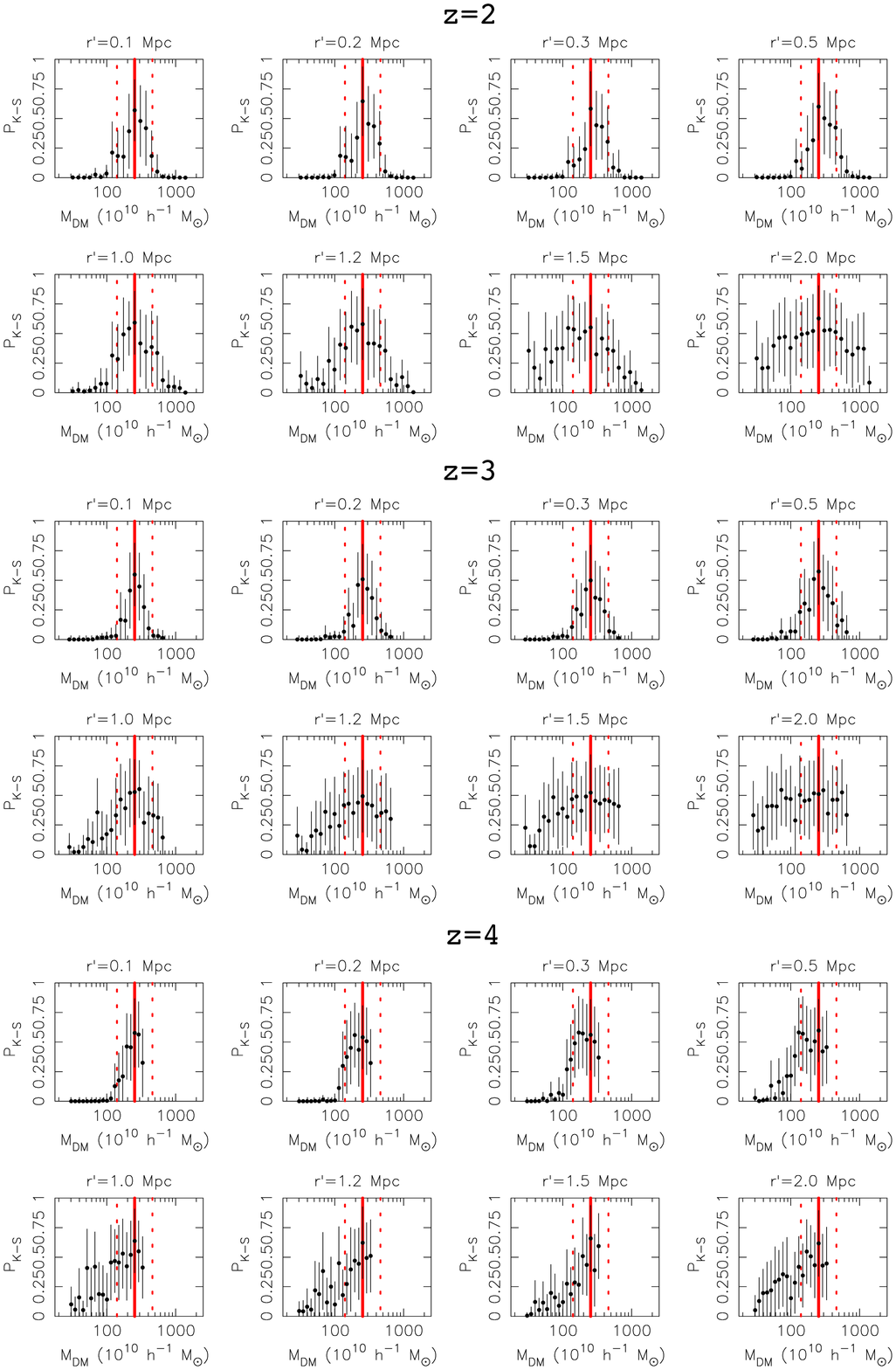}
\end{center}
\caption{
Kolmogorov-Smirnov P-value for the significance of the difference
between the true (\mbox{$M_{DM}=3.0\pm1.6~h^{-1}\times10^{12}$~\Msun}) and trial optical depth PDFs as a function of the quasar host dark matter halo mass range, at increasing apparent distances from the quasar.
Each trial mass range has the same logarithmic width as the true one
(indicated by the vertical red lines) and is labeled by the central
mass in logarithmic space.
In each case, the P-value is averaged over 50 samples of 100 typical quasar spectra at $z=4$ and the vertical error bars indicate the standard deviation.
At $z=4$, the K-S points do not appear to fall on the right side of the
true mass range simply because there are too few massive halos in the
simulation box to probe mass ranges much larger than the fiducial one.
}
\label{ks combined 100}
\end{figure*}

Figure \ref{ks combined 100} shows the results of this K-S analysis at
$z=2,~3,~\textrm{and}~4$.
Because the realized $P_{K-S}$ values vary with the random samples, we
show the mean value (dot) and the standard deviation (error bar) for
each mass, computed from ensembles of 50 sets of mock spectra.
At each redshift, the mass information appears mostly contained at
points $r'\lesssim1$ proper Mpc (note that this is still $\sim 20 r_{vir}$
-- see eq. \ref{virial radius}), although this conclusion is somewhat
dependent on the range of masses probed by our simulation.
Even if $P_{K-S}$ drops only marginally for masses differing from
$M_{DM}^{true}$ for the masses probed, the analysis would presumably
significantly rule out much smaller or much larger masses.
This is supported by Figure \ref{optical depth pdfs}, which shows that the
optical depth PDFs for halos located in random locations (effectively, $M_{DM}=0$) strongly differ from those of the massive halos resolved in the simulation.

\subsection{Test of the Halo Mass Likelihood}
\label{halo mass test}
\begin{figure}[ht]
\begin{center}
\includegraphics[width=0.475\textwidth]{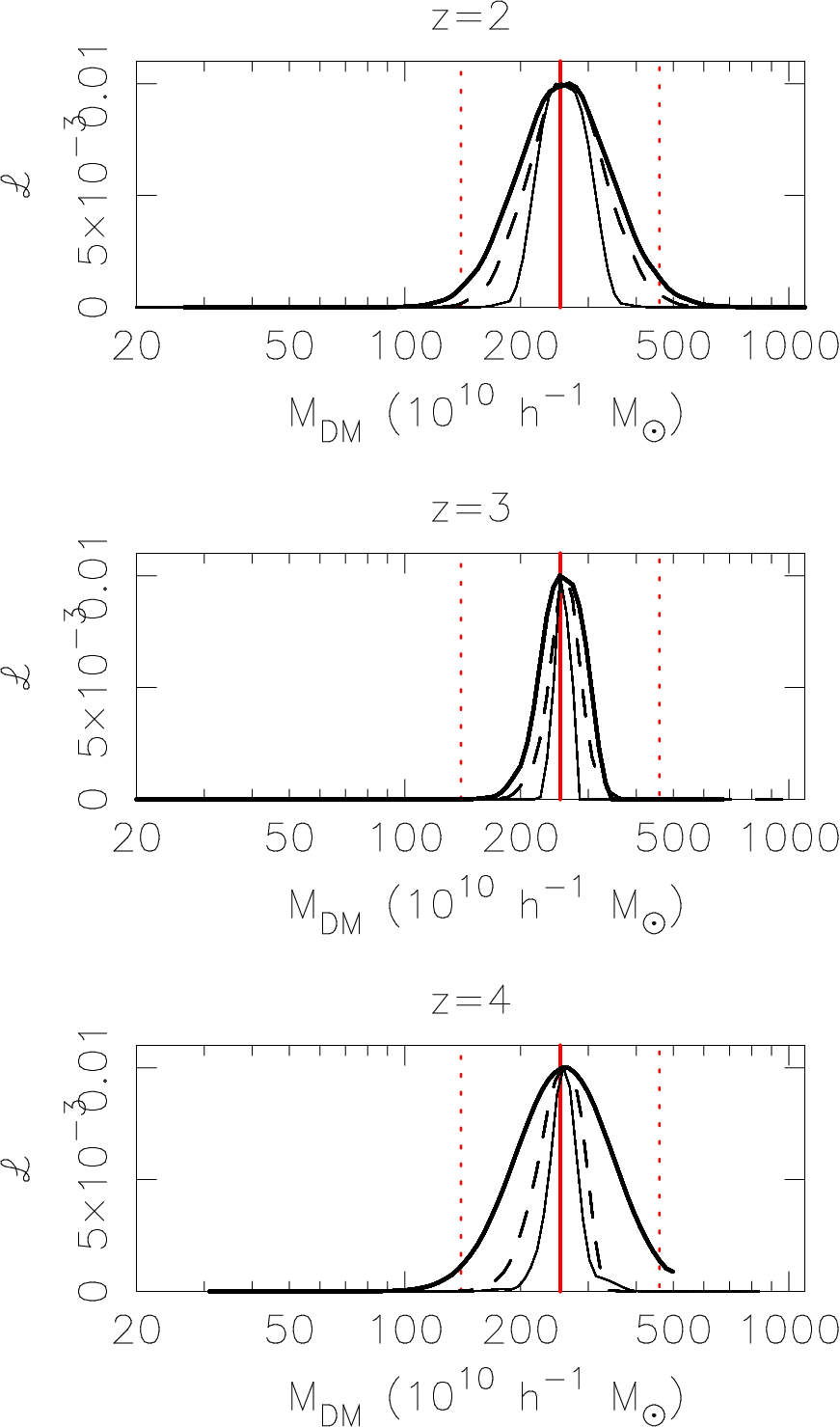}
\end{center}
\caption{
Likelihoods for $M_{DM}$ computed on samples of 100 mock spectra at $z=2,
3,~\textrm{and}~4$ (top, middle, and bottom) for quasars of typical luminosity lying at the centers of mass of dark matter halos of mass in the range \mbox{$3.0\pm1.6\times10^{12}~h^{-1}$ \Msun} (indicated by the vertical red lines).
\mbox{$\Gamma^{bkg}=10^{-12}$ s$^{-1}$} is assumed to be known
exactly.
The correlation length, $r_{corr}$, assumed in the computation of the likelihoods is 1.0 proper Mpc (thick solid curves), 0.5 proper Mpc (dashed curves), and 0.1 proper Mpc (thin solid cruves).
Better apparent constraints are obtained with smaller $r_{corr}$, as more data points are used.
However, the width of the estimated likelihoods is underestimated by an increasing factor as more correlated points are assumed independent (Appendix \ref{effect of correlations on likelihood}).
The correct mass range is recovered with a precision comparable to the
width of the true mass interval in the case $r_{corr}=1.0$ proper Mpc, in which case the correlations are small (Figure \ref{ln tau cf}).
The leftmost value on the horizontal axis corresponds to the
smallest-mass halos resolved in our simulation.
The $z=4$ likelihood is truncated at high masses simply because the
simulation box contains too few halos with mass much larger than the
fiducial range to evaluate the likelihood at those points.
The likelihoods on this figure were normalized to peak at 0.01.
}
\label{Mdm llhd}
\end{figure}

Figure \ref{Mdm llhd} shows likelihoods for $M_{DM}$ computed on
samples of 100 mock spectra for typical quasars lying in halos in the
fiducial mass range \mbox{$3.0\pm1.6\times10^{12}~h^{-1}$ \Msun}.
For these likelihood calculations, we use $(r_{min}, r_{max})=(0.1, 5.1)$ proper Mpc and vary the correlation length, $r_{corr}$, assumed in the computation of the likelihoods from 1.0 proper Mpc to 0.5 and 0.1 proper Mpc.
Better apparent constraints are obtained with smaller $r_{corr}$, as more data points are used.
However, the width of the estimated likelihoods is underestimated by an increasing factor as more correlated points are assumed independent (see Appendix \ref{effect of correlations on likelihood})).
The correct mass range is recovered with a precision comparable to the
width of the true mass interval in the case $r_{corr}=1.0$ proper Mpc, in which case the correlations are small (see Figure \ref{ln tau cf}).
This shows that, in principle, it is possible to measure quasar host halo masses from the proximity effect with a data set of modest size.
In fact, a data set of 100 spectra is two orders of magnitude smaller than
those used in state-of-the-art clustering analyses such as the
\cite{2005MNRAS.356..415C} one.
It is true, however, that the clustering analyses have relatively low
spectroscopic requirements.
As we discuss in \S \ref{practical considerations}, important challenges must be addressed before the level of precision estimated here can be achieved in practice.
In fact, realistic spectra have finite resolution and signal-to-noise ratio.
Moreover, the continuum level is not given and must be estimated, and systematic quasar redshifts are often not precisely known. 

\section{PRACTICAL CONSIDERATIONS}
\label{practical considerations}
Throughout this paper, we have optimistically assumed ideal data sets.
In particular, we have assumed spectra with infinite resolution and signal-to-noise ratio ($SN$), and perfect redshift determinations for the quasars.
In this section, we briefly discuss the complications associated with realistic data. 
As we will argue, \emph{if} the continuum flux and redshift of each quasar are known accurately, finite resolution and signal to noise can be accounted for and do not present serious problems.
However, our ability to estimate a quasar's continuum flux is limited by the resolution and noise in the spectra and accurate quasar systemic redshifts are challenging to obtain.
These difficulties are likely to be the most important in applying the method presented in this paper to measure quasar host halo masses to actual data.

\subsection{Finite Resolution and Signal-to-Noise Ratio}
Let us first suppose that the continuum flux and redshift of each quasar are given, and consider the effects of finite resolution and signal-to-noise ratio.

Finite resolution alters the optical depth PDF by smoothing out the small-scale fluctuations.
This can in principle be simply accounted for by smoothing the mock spectra from which the model PDF are constructed to the same resolution as the actual data.

A finite signal-to-noise ratio requires more spectra to detect intrinsic fluctuations in the IGM of a fixed amplitude.
The total variance on the pointwise transmission $F\equiv e^{-\tau}$ is a sum of the variance due to intrinsic IGM fluctuations, $\sigma_{F}^{2}$, and the variance in the observed flux due to the finite $SN$, $\sigma_{SN}^{2}\equiv1/SN^{2}$.
Thus, the error on the standard estimator for the mean $F$ estimated from $N$ spectra is
\begin{equation}
\sigma_{\langle F \rangle} =
\frac{ \sqrt{\sigma_{F}^{2} + \sigma_{SN}^{2}} }
{\sqrt{N}}
\end{equation}
The PDF shown on Figure \ref{optical depth pdfs} suggests that to distinguish between different halo masses requires measuring mean $F$ at the percent level.
Thus, one needs approximately
\begin{equation}
N=\frac{\sigma_{F}^{2} + 1/SN^{2}}{0.01^{2}}.
\end{equation}
spectra to distinguish between different realistic halo masses.
At $z\sim3$, typical intrinsic fluctuations have $\sigma_{F}\approx0.2$, thus requiring $SN\approx5$ to dominate over the noise.
This, by itself, is a modest requirement.
In that case, of order $10^{3}$ spectra are necessary to distinguish between different halo masses.
Of course, valuable mass information may also be encoded in the higher moments of the transmission, in particular in the incidence of large ($\sim1$) optical depths.
Several data points from each spectrum can also in principle be used (e.g., as in the likelihood method described in \S \ref{Gbkg ML}), alleviating the data requirements.

\subsection{Continuum Fitting}
An indirect, more severe effect of finite resolution and $SN$ is to make accurate continuum fitting more difficult. 
In the case of finite $SN$, the ``peaks'' in the observed spectra which are usually fitted to estimate the continua in general do not exactly correspond to a transmission of unity, owing to the noise contribution.
In the case of finite resolution, the peaks are smoothed and their height may therefore be underestimated. 
There is also a fundamental limit to how precisely quasar spectra can be continuum fitted, since peaks of unity transmission become increasingly rare, if extant at all, at high ($\gtrsim4$) redshifts. 
Continuum errors are of special concern because continuum fitting is a somewhat subjective procedure and they are likely to be affected by systematics to some extent.
It may thus be fruitful to devise more sophisticated methods which bypass continuum estimation, along the lines suggested by \cite{2006ApJ...638...27L} for the analysis of the matter power spectrum.

\subsection{Redshift Errors}
In practice, the uncertainties on quasar redshifts are substantial.
The redshifts determined from broad emission lines, which are affected by inflows and outflows of material, can differ a lot from the actual systemic redshift of the quasar \citep{1982ApJ...263...79G, 1992ApJS...79....1T, 2001AJ....122..549V, 2002AJ....124....1R}.
For example, \cite{2002AJ....124....1R} find that the broad C IV emission line has a median blueshift of 824 km s$^{-1}$ with respect to the narrow Mg II line, with a dispersion about the median of 511 km s$^{-1}$.
At $z=3$, this corresponds to a physical scale of a few Mpc, comparable to the radius which is substantially affected by the host halo of the quasar. 
Narrow emission lines, such as [O III] 5007 \AA~and Mg II 2798 \AA, which are associated with the host galaxy, can provide systemic redshifts to a precision $\approx50-300$ km s$^{-1}$ (J. Hennawi, private communication; see also Vanden Berk et al. 2001 and Richards et al. 2002a\nocite{2001AJ....122..549V, 2002AJ....124....1R}).
At $z\gtrsim2$, these narrow lines however require measurements extending in the near infrared.
The importance of accurate redshifts has been highlighted in the context of the transverse proximity effect by \cite{2006ApJ...651...61H} and \cite{2006astro.ph..6084H}.
Methods to obtain robust systemic quasar redshifts, like that developed by \cite{2006ApJ...651...61H} and \cite{2007AJ....133.2222S}, are therefore likely to be necessary for accurate proximity effect work.

\section{COMPARISON WITH OTHER WORK}
\label{previous work}
\cite{1995ApJ...448...17L} first estimated the bias in proximity
effect measurements of \Gbkg~arising from their overdense
environments.
They found that the bias could be up to a factor $\sim3$, consistent
with our results.
Their analytical analysis, very different from ours, was based on counting the \Lya~absorption lines Doppler shifted beyond the quasar redshift owing to gas infall.
\cite{2005MNRAS.361.1015R} also proposed using the proximity effect to
study the density structure around quasars.
They however incorrectly neglected the effects of gas infall toward
the quasars, which we have shown to be very important.
Instead of using dark matter halos from their simulation, they
simply modeled the density enhancement around quasars with an
analytic function of distance multiplying the density field.
This approach neglects the fact that the density fluctuations are not
necessarily linearly scaled as a quasar lying in a halo is approached, e.g. the fluctuations and their standard deviation need not be scaled by the same factor.
As Figure \ref{density profiles} shows, the ratio of $\langle \delta \rangle$ to the pointwise standard deviation of $\delta$ varies strongly with distance, violating the authors' assumption.
This approach also has the drawback of not relating the recovered density profiles to the mass of the quasar host halos.
\cite{2007MNRAS.tmp..238G} have applied the \cite{2005MNRAS.361.1015R} model on a sample of 45 quasars at $z>4$ and measured an average halo profile extending to $\gtrsim40$ proper Mpc, an order of magnitude farther than expected from simulations of structure formation (c.f., Figure \ref{density profiles}), casting doubt on the validity of their approach.
This anomalous result is likely an artifact of an inaccurate modeling of the ionizing effect of the quasars (the standard proximity effect), which is dominant and must be subtracted in order to measure the host halo mass.
For instance, these authors neglect redshift-space distortions altogether and their analytical model is partly heuristic and mathematically inexact.

\section{CONCLUSION}
\label{conclusion}
In this paper, we have shown that the \Lya~optical depth statistics in the
proximity regions of quasar spectra are significantly affected by the massive halos
which they are expected to occupy.
In general, the mean optical depth increases with halo mass.
We have then quantified the biases induced in line-of-sight proximity effect measurements
of the background photoionizing rate \Gbkg~that neglect the
effects of quasar host halos assuming idea data, i.e. that the quasar spectra have infinite resolution and signal-to-noise ratio and perfectly known continuum level and systemic redshift.
The local matter overdensity around and the infall of gas toward the quasars contribute approximately equally to the total upward bias.
At $z\approx3$, where most proximity effect measurements of
\Gbkg~have been made, the proximity effect \Gbkg~bias for host halo masses in the
range \mbox{$3.0\pm1.6~h^{-1}\times 10^{12}$ \Msun}, as inferred by
\cite{2005MNRAS.356..415C} from clustering measurements, could
be $\approx2.5$, enough to bring in agreement the existing proximity effect
and flux decrement measurements.
The existing proximity effect measurements were however made on data of finite quality, with continua and redshifts estimated from the data.
These observational difficulties may also affect the validity of those proximity effect measurements, beyond the effects investigated here.
The fact that quasars lying in overdense regions of the Universe introduces very significant biases is nonetheless a robust conclusion of our study, and should be taken into account in future analyses.

The clustering of galaxies and other AGN around proximity effect
quasars has a small effect on the local magnitude of the background
ionizing flux, at least at $z\lesssim3$, and therefore is not expected
to significantly bias \Gbkg~measurements.

By constructing optical depth PDFs by Monte
Carlo from realistic mock spectra with quasars in host halos of
prescribed mass, one can perform a likelihood analysis for \Gbkg~which
is unbiased.
However, this method is sensitive to the absolute level of
\Lya~absorption, both close to and far from the quasars.
In the limit where only points far from the quasars are considered in
the analysis, it essentially reduces to the flux decrement method.
As such, the method does not possess the main advantage of ``true''
proximity effect analyses (which are only sensitive to the \emph{change} in
absorption statistics near the quasars) that they are relatively free
of cosmological parameter assumptions.
In particular, this method requires knowledge of $\Omega_{b}$ and so
a \Gbkg~value obtained with it cannot be used in conjunction with the
flux decrement to constrain the baryon density.

Of perhaps greatest interest, we have demonstrated how, given a
measurement of \Gbkg (e.g., from the flux decrement), the proximity effect analysis can in principle be inverted to probe the environments of quasars.
In particular, we have shown that the masses of dark matter halos
hosting quasars could be measured using the optical depth
statistics in the proximity regions of quasars.

As mentioned, we have in this work side-stepped a number of observational issues.
For instance, we have assumed that the continuum flux and redshift of
each quasar were known exactly; in practice, these are plagued by
uncertainties.
Moreover, real spectra have finite resolution and signal-to-noise
ratio.
From our discussion of these complications, we have concluded that accurate continuum fitting and quasar redshift determination are likely to be the most important challenges to using our method to measure halo masses.
Further work in developing techniques to either improve or bypass these measurements would thus be highly desirable.
Our theoretical picture is also simplified in some respects.
For instance, we have assumed that the low-density \cite{1997MNRAS.292...27H} equation of state for the IGM holds all the way to the centers of the quasar host halos, which may not be the case.
In particular, if helium is doubly-ionized by quasars at $z \sim 3$ (e.g. \citealt{1999ApJ...514..648M}),
the temperature should be enhanced close to quasar sources.
Whether such thermal effects are important in the vicinity of quasars
could be investigated by studying how the small-scale power spectrum in the
\Lya~forest varies as the quasars are approached, as was proposed by
\cite{2002ApJ...564..153Z} for the general forest.

\acknowledgements
We thank Suvendra Dutta for assistance with the simulations
used in this work.
We also acknowledge enlightening discussions with Joe Hennawi, Jason Prochaska, and Scott Burles on practical issues in the analysis of quasar spectra.
CAFG is supported by a NSERC Postgraduate Fellowship and Harvard
University grants.
This work was supported in part by NSF grants ACI
96-19019, AST 00-71019, AST 02-06299, AST 03-07690, and AST 05-06556, and NASA ATP
grants NAG5-12140, NAG5-13292, NAG5-13381, and NNG-05GJ40G.
Further support was provided by the David and Lucile Packard, the Alfred P. Sloan, and the John D. and Catherine T. MacArthur Foundations.

\appendix
\section{A. EXISTING FLUX DECREMENT AND PROXIMITY EFFECT MEASUREMENTS OF THE BACKGROUND PHOTOIONIZING FLUX}
\label{measurements appendix}
In this appendix, we tabulate the existing measurements of \Gbkg~from the flux decrement and proximity effect methods.
This serves as a quantitative complement to Figure \ref{existing measurements figure}.

\begin{center}
\begin{deluxetable}{cccc}
\tablecaption{Existing Flux Decrement and Proximity Effect Measurements of the Background Photoionizing Flux\label{existing measurements table}}
\tablewidth{0pt}

\tablehead{
\multicolumn{4}{c}{Flux Decrement} \\
\hline
\colhead{Redshift} & \colhead{$J_{-21}$} & \colhead{\Gbkg} & \colhead{Reference} \\
\colhead{} & \colhead{ergs s$^{-1}$ cm$^{-2}$ Hz$^{-1}$ sr$^{-1}$} & \colhead{$10^{-12}$ s$^{-1}$} & \colhead{}}

\startdata
1.95 & & 1.33 & \cite{2005MNRAS.361...70J} \\

3 & & $1.3\pm0.1$ & \cite{2005MNRAS.360.1373K} \\

2 & & $1.3^{+0.8}_{-0.5}$ & \cite{2005MNRAS.357.1178B} \\
3 & & $0.9\pm0.3$         & $\ldots$ \\
4 & & $1.0^{+0.5}_{-0.3}$ & $\ldots$ \\

1.9 & & $1.44\pm0.11$ & \cite{2004ApJ...617....1T} \\

2.75 & & $0.86^{+0.36}_{-0.24}$  & \cite{2004MNRAS.350.1107M} \\
3.0  & & $0.88^{+0.14}_{-0.12}$  & $\dots$ \\
3.89 & & $0.68^{+0.08}_{-0.07}$  & $\dots$ \\
4.0  & & $0.43^{+0.06}_{-0.05}$  & $\dots$ \\
5.0  & & $0.31^{+0.07}_{-0.09}$  & $\dots$ \\
5.5  & & $0.37^{+0.06}_{-0.05}$  & $\dots$ \\
6.0  & & $<0.14$  & $\dots$ \\ 

2.4  & & $0.698\pm0.096$ & \cite{2001ApJ...549L..11M} \\
3    & & $0.518\pm0.083$ & $\dots$ \\
3.9  & & $0.380\pm0.04$  & $\dots$ \\
4.5  & & $0.21\pm0.04$   & $\dots$ \\
4.93 & & $0.13\pm0.03$   & $\dots$ \\
5.2  & & $0.16\pm0.04$   & $\dots$ \\

4.72 & $\gg0.04$\tablenotemark{a} & $\gg$0.129  & \cite{1999ApJ...525L...5S} \\

$[1.5, 2.5]$ & & 0.890 & \cite{1997ApJ...489....7R}\tablenotemark{b} \\
$[2.5, 3.5]$ & & 0.698 & $\ldots$ \\
$[3.5, 4.5]$ & & 0.618 & $\ldots$ \\

\hline
\multicolumn{4}{c}{Proximity Effect} \\
\hline

$[1.7, 3.8]$ & $0.7^{+0.34}_{-0.44}$ & $1.9^{+1.2}_{-1.0}$\tablenotemark{c,d} & \cite{2000ApJS..130...67S} \\

$[2.0, 4.5]$ & $1.0^{+0.5}_{-0.3}$ & $2.6^{+1.3}_{-0.8}$ & \cite{1997MNRAS.284..552C} \\ 

$[1.7, 3.7]$ & 0.6 & 1.6\tablenotemark{*} & \cite{1996MNRAS.280..767S} \\

$[1.7, 4.1]$ & $0.5\pm0.1$ & $1.3\pm0.3$\tablenotemark{*} & \cite{1996ApJ...466...46G} \\

3.66 & $0.5$\tablenotemark{e} & $1.3$\tablenotemark{*} & \cite{1995MNRAS.273.1016C} \\

$\approx4.2$ & $\approx0.1-0.3$\tablenotemark{f} & $0.3-0.8$\tablenotemark{*} & \cite{1994ApJ...428..574W} \\

$\approx0.5$ & $\approx0.006$\tablenotemark{g} & $0.02$\tablenotemark{*} & \cite{1993ApJ...413L..63K} \\

$[1.7, 3.8]$ & $1^{+2}_{-0.7}$ & $2.6^{+4.2}_{-1.8}$\tablenotemark{*} & \cite{1991ApJ...367...19L} \\

$[1.7, 3.8]$ & $1^{+3.2}_{-0.7}$ & $2.6^{+8.3}_{-1.8}$\tablenotemark{*} & \cite{1988ApJ...327..570B} \\

3.75 & $\gtrsim3$\tablenotemark{h} & $\gtrsim7.8$\tablenotemark{*} & \cite{1987ApJ...319..709C} \\
\enddata

\tablenotetext{a}{Assuming a spectral index of 0.7 for the background flux.}
\tablenotetext{b}{For their $\Lambda$CDM cosmology.}
\tablenotetext{c}{The authors claim that the presence of lines on the
saturated part of the curve of growth could cause their estimate to be
overestimated by a factor 2$-$3.}
\tablenotetext{d}{Contrary to other proximity effect analyses, this value does not assume a spectral index for the background; the authors repeated their analysis solving directly for \Gbkg.}
\tablenotetext{e}{From a single quasar, QSO 0055$-$269.}
\tablenotetext{f}{From a single quasar, QSO BR 1033$-$0327.}
\tablenotetext{g}{The uncertainties are large; at the $1\sigma$ could
be lower by a factor of 3 or higher by a factor of 6.}
\tablenotetext{h}{From a single quasar, PKS 2000$-$330.}
\tablenotetext{*}{Calculated from $J_{\nu_{Ly}}$ assuming a spectral index for the background flux equal to the typical value for radio-quiet quasars, $\alpha=1.57$ \citep{2002ApJ...565..773T}. These authors assumed that the background flux had the same spectral index as the quasars shortward of the Lyman limit in their analyses, from which they inferred $J_{\nu_{Ly}}$.}

\end{deluxetable}
\end{center}

\section{B. ACCURATE EXPRESSIONS}
\label{accurate expressions}
In some of our calculations, we have used approximate expressions for
the physical properties of the IGM in order to improve computational
efficiency.
These approximations are not expected to have any significant effect
on our results, as they were applied consistently.
However, when the methods presented in this
work are to be applied to real spectra, it may be important to use more accurate expressions,
in order to be consistent with the physical processes as they
occur in the Universe.
Some lengthy formulae were also omitted in the main body in order to improve readability.
In this Appendix, based on that of \cite{1997MNRAS.292...27H}, we collect  explicit accurate expressions missing in the text.

Hydrogen photoionization cross section:
\begin{equation}
\sigma(E)=
5.475 \times 10^{-14}\textrm{ cm}^{2} (E/0.4298\textrm{ eV} - 1)^{2} 
\frac{(E/0.4298\textrm{ eV})^{-4.018}}
{(1 + \sqrt{E/14.13\textrm{ eV}})^{2.963}}
\end{equation}
(Verner et al. 1996\nocite{1996ApJ...465..487V}; accurate to 10\% from the ionization thresholds to \mbox{5 keV}).

Hydrogen recombination coefficient:
\begin{equation}
R(T)=
1.269 \times 10^{-13}\textrm{ cm}^{-3}\textrm{ s}^{-1}
\frac{(2 \times 157,807\textrm{ K} / T)^{1.503}}
{[1.0 + (2 \times 157,807\textrm{ K} / 0.522T)^{0.470}]^{1.923}}
\end{equation}
(fit by Hui \& Gnedin 1997\nocite{1997MNRAS.292...27H} to Ferland et al. 1992\nocite{1992ApJ...387...95F} data; accurate to 2\%
from 3 to \mbox{10$^{9}$ K}).

$T_{0}$ coefficient in IGM equation of state \ref{igm eos}:
\begin{equation}
T_{0}^{1.7}
=
\left(
\frac{a_{reion}^{2}}
{a^{2}}
T_{reion}
\right)^{1.7}
+
\frac{1.7}
{1.9}
a^{-[3/2 + (a - 0.25)/a]}
\left(
1 -
\frac{a_{reion}^{1.9}}
{a}
\right)
B,
\end{equation}
where
\begin{equation}
B = 
\frac{\rho_{b}}
{H_{0} m_{p} \sqrt{\Omega_{m}}}
T_{reion}
R(T_{reion}),
\end{equation}
\mbox{$T_{reion}=24,000$ K}, and we take $z_{reion}=1/(1 + a_{reion})=10$.

IGM equation of state exponent $\beta$:
\begin{equation}
\beta = 
\frac{1}
{1.7}
\left[
1 -
\left(
\frac{a^{2}_{reion} T_{reion}}
{a^{2} T_{0}}
\right)^{1.7}
\frac{a_{reion}}
{a}
\right]
+
\left(
\frac{2}
{3}
-
\frac{1}
{1.7}
\right)
(a^{2} T_{0})^{-1.7}
C,
\end{equation}
where
\begin{equation}
C =
\frac{1.7}
{1.9}
B
\left\{
\frac{a^{1.9}}
{2.9}
\left[
1 - 
\left(
\frac{a_{reion}}
{a}
\right)^{2.9}
\right]
-
a_{reion}^{1.9}
\left(
1 -
\frac{a_{reion}}
{a}
\right)
\right\}
+
(a_{reion}^{2} T_{reion})^{1.7}
\left(
1 -
\frac{a_{reion}}
{a}
\right).
\end{equation}

\section{C. THE EDGEWORTH EXPANSION}
\label{edgeworth expansion}
The Edgeworth expansion \citep[e.g.,][]{1995ApJ...442...39J, 1998A&AS..130..193B} is an expansion of a nearly Gaussian distribution in terms of its moments.
Since the \Lya~optical depth statistics are nearly lognormal, their
PDF is well approximated by an Edgeworth expansion in the first four
moments:
\begin{equation}
\label{ln tau PDF edgeworth expansion}
f_{\ln{\widehat{\tau}}}(\ln{\tau}) \approx
\frac{1}{\sigma_{\ln{\tau}}}
\left|
\left\{
1 
+ \frac{1}{3!}
\textrm{skew}(\ln{\tau}) H_{3}(\eta)
+ \frac{1}{4!}
\textrm{kurt}(\ln{\tau}) H_{4}(\eta)
+ \frac{10}{6!}
[\textrm{skew}(\ln{\tau})]^{2} H_{6}(\eta)
\right\}
\phi(\eta) 
\right|
_{\eta = \frac{\ln{\tau} - \langle \ln{\tau} \rangle}{\sigma_{\ln{\tau}}}},
\end{equation}
where $\textrm{skew}(\ln{\tau})$ and $\textrm{kurt}(\ln{\tau})$ are
the skewness and kurtosis of $f_{\ln{\widehat{\tau}}}$, respectively, the $H_{n}(x)$
are Hermite polynomials \citep[e.g.,][]{1965hmfw.book.....A}, and
\begin{equation}
\phi(\eta) = 
\frac{1}{2 \pi}
\exp{
\left(
-\frac{\eta^2}{2}
\right)
}
\end{equation}
is a standard Gaussian PDF.
Unbiased estimators of the skewness and kurtosis are given by
\begin{equation}
\label{skew estimator}
\widehat{\textrm{skew}(\ln{\tau)}} = 
\frac{N\sqrt{N-1}}
{N - 2}
\frac{\sum_{i=1}^{N}(\ln{\tau_{i}} - \widehat{\langle \ln{\tau} \rangle})^{3}}
{[\sum^{N}_{i=1} (\ln{\tau_{i}} - \widehat{\langle \ln{\tau}
\rangle})^{2}
]^{3/2}}
\end{equation}
and
\begin{equation}
\label{kurt estimator}
\widehat{\textrm{kurt}(\ln{\tau)}} = 
\frac{(N+1)N(N-1)}
{(N-2)(N-3)}
\frac{
\sum_{i=1}^{N} (\ln{\tau_{i}} - \widehat{\langle \ln{\tau} \rangle})^{4}
}
{
[\sum_{i=1}^{N} (\ln{\tau_{i}} - \widehat{\langle \ln{\tau} \rangle})^{2}]^{2}
}
- 3\frac{(N-1)^{2}}
{(N-2)(N-3)}.
\end{equation}
The Edgeworth expansion is used to approximate PDFs of $\ln{\tau}$ in
sections \ref{unbiased llhd} and \ref{halo mass}.

\section{D. EFFECT OF CORRELATIONS ON THE LIKELIHOOD FUNCTION}
\label{effect of correlations on likelihood}
In this section, we present a toy model to understand the effect of correlations on the likelihood function.
Specifically, we seek to quantity the error that is made if neighboring data points are correlated, but a likelihood estimator assuming independence is used.
For example, in this paper we pretended that points separated by a ``correlation length'' were perfectly independent (e.g., in \S \ref{Gbkg ML}).
We would like to know how well our estimate of the likelihood obtained in this way represents the true likelihood. \\ \\
Consider $N$ independent pairs of Gaussian random variables $X_{i,1}$ and $X_{i,2}$, each with mean $\mu$ and variance $\sigma^{2}$.
The Gaussian functional form is well-motivated, since $\ln{\tau}$ is approximately normally distributed.
$X_{i,1}$ and $X_{i,2}$ are correlated with one another, with covariance matrix
\begin{equation}
\Sigma = 
\left( \begin{array}{cc}
\sigma^{2} & \zeta      \\
\zeta      & \sigma^{2} \\
\end{array} \right),
\end{equation}
so that their joint PDF is given by
\begin{equation}
\label{bivariate gaussian}
f(\mathbf{x})=\frac{1}{2 \pi |\Sigma|^{1/2}}
\exp{ 
\left[ 
(\bf{x - \mu}) \Sigma^{-1} (\bf{x - \mu})\transpose
\right]
},
\end{equation}
where $\bf{x}$$\equiv(x_{1}, x_{2})$.
Let $\mathcal{L}^{i,est}$ be the likelihood for $\mu$ estimated from $X_{i,1}$ and $X_{i,2}$, ignoring the correlation between the two random variables, and $\mathcal{L}^{i,true}$ be the ``true'' likelihood.
Then:
\begin{equation}
\ln{ \mathcal{L}^{i,est} } = 
-\ln{[2 \pi \sigma]} 
- \frac{1}{2 \sigma^{2}}
\left[
(x_{1} - \mu)^{2} + (x_{2} - \mu)^{2}
\right]
\end{equation}
and
\begin{equation}
\ln{\mathcal{L}^{i,true}} = 
-\ln{[2 \pi \sigma (1 - \epsilon^{2})^{1/2}]} 
- \frac{1}{2 \sigma^{2} (1 - \epsilon^{2})}
\left[
(x_{1} - \mu)^{2} - 2\epsilon(x_{1} - \mu)(x_{2} - \mu) + (x_{2} - \mu)^{2}
\right],
\end{equation}
where $\epsilon \equiv \zeta/\sigma^{2}$.
Given that $(X_{i,1}, X_{i,2})$ is distributed as in equation \ref{bivariate gaussian}, we may compute the expectation values of $\ln{\mathcal{L}^{i,est}}$ and $\ln{\mathcal{L}^{i,true}}$:
\begin{equation}
\langle \ln{\mathcal{L}^{i,est}} \rangle =
- \ln{[2 \pi \sigma^{2}]}
- 
\left[
1 + 
\left( 
\frac{\Delta \mu}{\sigma}
\right)^{2}
\right]
\end{equation}
and
\begin{equation}
\langle \ln{\mathcal{L}^{i,true}} \rangle =
- \ln{[2 \pi \sigma^{2}(1 - \epsilon^{2})^{1/2}]}
- 
\frac{1}{1-\epsilon^{2}}
\left[
1 + 
\left( 
\frac{\Delta \mu}{\sigma}
\right)^{2}
\right]
+
\frac{\epsilon}{1 - \epsilon^{2}}
\left[
\epsilon + \left( \frac{\Delta \mu}{\sigma} \right)^{2}
\right],
\end{equation}
where $\Delta \mu$ is the difference between the true mean and the value at which the likelihood is evaluated.
Since $\langle \ln{\mathcal{L}^{i,est}} \rangle$ is maximum for $\Delta \mu=0$, the maximum likelihood estimate for $\mu$ is unbiased, even if the estimator ignores the correlations between the random variables.
An estimate for the width of $\langle \ln{\mathcal{L}} \rangle$ is given by the curvature at the maximum likelihood value, $\widetilde{\sigma} (\mathcal{L}) \equiv -[\sqrt{\partial^{2} \langle \ln{\mathcal{L}} \rangle / \partial (\Delta \mu)^{2}}]^{-1}$.
This expression is exactly equal to the standard deviation of the likelihood function when the latter is Gaussian.
Explicitly,
\begin{equation}
\widetilde{\sigma} (\mathcal{L}^{i,est})
= \frac{\sigma}{\sqrt{2}}
\textrm{~~~~~and~~~~~}
\widetilde{\sigma} (\mathcal{L}^{i,true})
= \sqrt{1 + \epsilon} \frac{\sigma}{\sqrt{2}}.
\end{equation}
Let us now consider the total likelihood for $\mu$, as estimated from the $N$ pairs $(X_{i,1}, X_{i,2})$.
By independence,
\begin{equation}
\mathcal{L}^{tot} = 
\prod_{i=1}^{N} \mathcal{L}^{i},
\end{equation}
so that
\begin{equation}
\ln{\mathcal{L}^{tot}} = 
\sum_{i=1}^{N} \ln{\mathcal{L}^{i}}.
\end{equation}
Therefore, 
\begin{equation}
\langle \ln{\mathcal{L}^{tot}} \rangle = N \langle \ln{\mathcal{L}^{i}} \rangle,
\end{equation}
and
\begin{equation}
\widetilde{\sigma} (\mathcal{L}^{i,est})
= \frac{\sigma}{\sqrt{2N}}
\textrm{~~~~~and~~~~~}
\widetilde{\sigma} (\mathcal{L}^{i,true})
= \sqrt{1 + \epsilon} \frac{\sigma}{\sqrt{2N}}.
\end{equation}
Thus, the width of the estimated likelihood, when correlations are ignored, is underestimated by a factor $1/\sqrt{1 + \epsilon}$ ($\approx1 - \epsilon/2$ for $\epsilon \ll 1$).
In particular, the error on the width of the likelihood is independent of the number of independent pairs of correlated data.
In the limit $\epsilon \to 0$ (no correlation), $\widetilde{\sigma} (\mathcal{L}^{i,est}) \to \widetilde{\sigma} (\mathcal{L}^{i,true})$, and the standard result for the error on the mean is recovered.
In the limit $\epsilon \to 1$ (the two data points from each pair are the same), $\widetilde{\sigma} (\mathcal{L}^{i,est}) \to \widetilde{\sigma} (\mathcal{L}^{i,true})/\sqrt{2}$, again as expected, since there are half as many data independent data points as the number of points from which the mean is estimated.

\bibliography{references}

\end{document}